\documentclass[usenatbib]{mnras}

\usepackage{amssymb}	

\usepackage{newtxtext,newtxmath}

\usepackage[T1]{fontenc}
\usepackage[normalem]{ulem}

\DeclareRobustCommand{\VAN}[3]{#2}
\let\VANthebibliography\thebibliography
\def\thebibliography{\DeclareRobustCommand{\VAN}[3]{##3}\VANthebibliography}

\usepackage{graphicx}	
\usepackage{amsmath}	

\usepackage{fontawesome5}
\usepackage{color}
\usepackage{xcolor}

\newcommand{\Rfivehc}{R_{\rm 500c}}
\newcommand{\Mfivehc}{M_{\rm 500c}}
\newcommand{\Rtwohc}{R_{\rm 200c}}
\newcommand{\Mtwohc}{M_{\rm 200c}}
\newcommand{\Rtwohm}{R_{\rm 200m}}
\newcommand{\Mtwohm}{M_{\rm 200m}}

\newcommand{\MeanY}{\langle y \rangle}

\newcommand{\Rshde}{R_{\rm sh,\,de}}
\newcommand{\Rshacc}{R_{\rm sh,\,acc}}
\newcommand{\Rsp}{R_{\rm sp}}

\newcommand{\eg}{{\sl e.g.}, }        

\newcommand{\msol}{\ensuremath{\, {\rm M}_\odot}}    
\newcommand{\msun}{\ensuremath{\, {\rm M}_\odot}} 
         
\newcommand{\mpc}{\ensuremath{\, {\rm Mpc}}}         
\newcommand{\gpc}{\ensuremath{\, {\rm Gpc}}}
     
\newcommand{\dln}{\ensuremath{{\rm d \ln}}}

\definecolor{orcidlogocol}{HTML}{A6CE39}
\definecolor{purple}{RGB}{128, 0, 128}

\newcommand{\OrcidID}[1]{ \href[urlcolor = red]{https://orcid.org/#1}{\textcolor{lightgray}{\faOrcid}}}
\newcommand{\OrcidIDName}[2]{\href{https://orcid.org/#1}{#2}}

\newcommand*{\vcenteredhbox}[1]{\begingroup
\setbox0=\hbox{#1}\parbox{\wd0}{\box0}\endgroup}

\title[Shocks in SPT-SZ Clusters]{Shocks in the Stacked Sunyaev-Zel’dovich Profiles of Clusters II: Measurements from SPT-SZ + {\it Planck} Compton-$y$ Map}

\author[Dhayaa Anbajagane et. al]{\OrcidIDName{0000-0003-3312-909X}{D. Anbajagane}\thanks{Corresponding author email: dhayaa@uchicago.edu}(\vcenteredhbox{\includegraphics[height=1.2\fontcharht\font`\B]{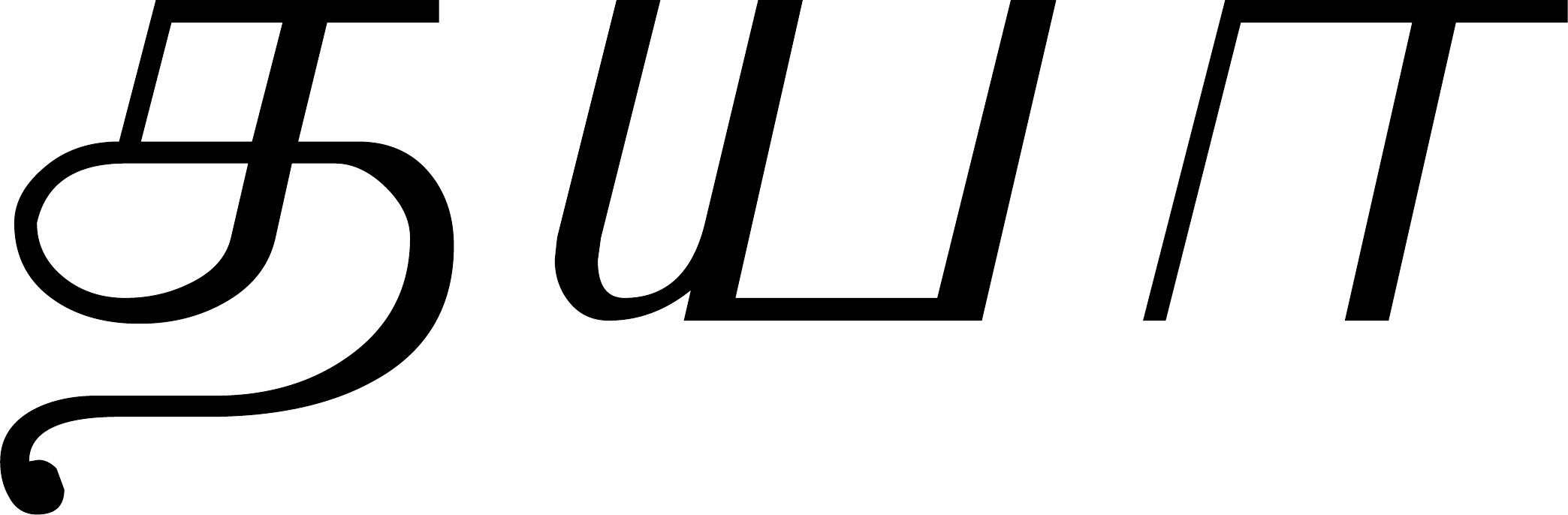}}),$^{1,\,2}$
\OrcidIDName{0000-0002-7887-0896}{C. Chang},$^{1,\,2}$
\OrcidIDName{0000-0002-8220-3973}{B. Jain},$^{3}$ 
\OrcidIDName{0000-0002-0298-4432}{S. Adhikari},$^{1,\,2}$ 
\OrcidIDName{0000-0002-6836-3196}{E.~J.~Baxter},$^{4}$ \newauthor
\OrcidIDName{0000-0002-5108-6823}{B.~A.~Benson},$^{5,\,2,\,1}$
\OrcidIDName{0000-0001-7665-5079}{L. E. Bleem},$^{6,\,2}$
\OrcidIDName{0000-0002-4900-805X}{S. Bocquet},$^{7}$
\OrcidIDName{0000-0002-2238-2105}{M. S. Calzadilla},$^{8}$
\OrcidIDName{0000-0002-2044-7665}{J.~E.~Carlstrom},$^{2,\,1,\,9,\,6,\,10}$\newauthor 
\OrcidIDName{0000-0002-6311-0448}{C.~L.~Chang},$^{6,\,2,\,1}$
\OrcidIDName{0000-0001-8241-7704}{R.~Chown},$^{11}$
\OrcidIDName{0000-0001-9000-5013}{T. M. Crawford},$^{1,\,2}$
A.~T.~Crites,$^{12,\,2,\,1}$
W. Cui,$^{13}$ 
T.~de~Haan,$^{14,\,15}$\newauthor
\OrcidIDName{0000-0003-3586-4485}{L. Di Mascolo},$^{16,\,17,\,18}$
M.~A.~Dobbs,$^{19,\,20}$
\OrcidIDName{0000-0002-5370-6651}{W.~B.~Everett},$^{21}$
\OrcidIDName{0000-0001-7874-0445}{E.~M.~George},$^{22,\,15}$
S. Grandis,$^{7,\,23}$ \newauthor
\OrcidIDName{0000-0003-2606-9340}{N.~W.~Halverson},$^{21,\,24}$
G.~P.~Holder,$^{25,\,26}$ 
W.~L.~Holzapfel,$^{15}$
J.~D.~Hrubes,$^{27}$
\OrcidIDName{0000-0003-3106-3218}{A.~T.~Lee},$^{15,\,29}$\newauthor 
D.~Luong-Van,$^{27}$ 
M. A. McDonald,$^{8}$ 
J.~J.~McMahon,$^{2,\,1,\,9,\,10}$
S.~S.~Meyer,$^{2,\,1,\,9,\,10}$
M.~Millea,$^{15}$\newauthor
L.~M.~Mocanu,$^{2,\,1}$
\OrcidIDName{0000-0002-6875-2087}{J.~J.~Mohr},$^{7,\,23,\,30}$
T.~Natoli,$^{2,\,1}$
Y.~Omori,$^{2,\,1,\,31,\,32}$
S.~Padin,$^{33,\,2,\,1}$
C.~Pryke,$^{34}$\newauthor 
\OrcidIDName{0000-0003-2226-9169}{C.~L.~Reichardt},$^{36}$
J.~E.~Ruhl,$^{36}$
A. Saro$^{16,\,17,\,18}$
K.~K.~Schaffer,$^{37,\,2,\,10}$
\OrcidIDName{0000-0002-2757-1423}{E.~Shirokoff},$^{2,\,1,\,15}$\newauthor 
Z.~Staniszewski,$^{38,\,36}$
\OrcidIDName{0000-0002-2718-9996}{A.~A.~Stark},$^{39}$
\OrcidIDName{0000-0001-7192-3871}{J.~D.~Vieira},$^{25,\,26}$
R.~Williamson$^{38,\,2,\,1}$\\
\\
$^{1}$ Department of Astronomy and Astrophysics, University of Chicago, Chicago, IL 60637, USA\\
$^{2}$ Kavli Institute for Cosmological Physics, University of Chicago, Chicago, IL 60637, USA\\
$^{3}$ Department of Physics and Astronomy, Center for Particle Cosmology, University of Pennsylvania, Philadelphia, PA 19104, USA\\
$^{4}$ Institute for Astronomy, University of Hawai’i, 2680 Woodlawn Drive, Honolulu, HI 96822, USA\\
$^{5}$ Fermi National Accelerator Laboratory, MS209, P.O. Box 500, Batavia, IL 60510\\
$^{6}$ High Energy Physics Division, Argonne National Laboratory, Argonne, IL, USA 60439\\
$^{7}$ Faculty of Physics, Ludwig-Maximilians-Universit\"at, Scheinerstr. 1, 81679 Munich, Germany\\
$^{8}$ Kavli Institute for Astrophysics and Space Research, Massachusetts Institute of Technology, Cambridge, MA 02139, USA\\
$^{9}$ Department of Physics, University of Chicago, Chicago, IL, USA 60637\\
$^{10}$ Enrico Fermi Institute, University of Chicago, Chicago, IL, USA 60637\\
$^{11}$ Department of Physics and Astronomy, McMaster University, 1280 Main St. W., Hamilton, ON L8S 4L8, Canada\\
$^{12}$ Department of Astronomy \& Astrophysics, University of Toronto, 50 St George St, Toronto, ON, M5S 3H4, Canada\\
$^{13}$ Institute for Astronomy, University of Edinburgh, Royal Observatory, EH9 3HJ Edinburgh, United Kingdom\\
$^{14}$ High Energy Accelerator Research Organization (KEK), Tsukuba, Ibaraki 305-0801, Japan\\
$^{15}$ Department of Physics, University of California, Berkeley, CA, USA 94720\\
$^{16}$ Astronomy Unit, Department of Physics, University of Trieste, via Tiepolo 11, Trieste 34131, Italy\\
$^{17}$ INAF - Osservatorio Astronomico di Trieste, via Tiepolo 11, Trieste 34131, Italy\\
$^{18}$ IFPU - Institute for Fundamental Physics of the Universe, Via Beirut 2, 34014 Trieste, Italy\\
$^{19}$ Department of Physics and McGill Space Institute, McGill University, Montreal, Quebec H3A 2T8, Canada\\
$^{20}$ Canadian Institute for Advanced Research, CIFAR Program in Cosmology and Gravity, Toronto, ON, M5G 1Z8, Canada\\
$^{21}$ Center for Astrophysics and Space Astronomy, Department of Astrophysical and Planetary Sciences, University of Colorado, Boulder, CO, 80309\\
$^{22}$ European Southern Observatory, Karl-Schwarzschild-Stra{\ss}e 2, 85748 Garching, Germany\\
$^{23}$ Excellence Cluster ORIGINS, Boltzmannstr. 2, 85748 Garching, Germany\\
$^{24}$ Department of Physics, University of Colorado, Boulder, CO, 80309\\
$^{25}$ Astronomy Department, University of Illinois at Urbana-Champaign, 1002 W. Green Street, Urbana, IL 61801, USA\\
$^{26}$ Department of Physics, University of Illinois Urbana-Champaign, 1110 W. Green Street, Urbana, IL 61801, USA\\
$^{27}$ University of Chicago, Chicago, IL, USA 60637\\
$^{29}$ Physics Division, Lawrence Berkeley National Laboratory, Berkeley, CA, USA 94720\\
$^{30}$ Max-Planck-Institut f\"{u}r extraterrestrische Physik, 85748 Garching, Germany\\
$^{31}$ Kavli Institute for Particle Astrophysics and Cosmology, Stanford University, 452 Lomita Mall, Stanford, CA 94305\\
$^{32}$ Dept. of Physics, Stanford University, 382 Via Pueblo Mall, Stanford, CA 94305\\
$^{33}$ California Institute of Technology, Pasadena, CA, USA 91125\\
$^{34}$ Department of Physics, University of Minnesota, Minneapolis, MN, USA 55455\\
$^{35}$ School of Physics, University of Melbourne, Parkville, VIC 3010, Australia\\
$^{36}$ Physics Department, Center for Education and Research in Cosmology and Astrophysics, Case Western Reserve University,Cleveland, OH, USA 44106\\
$^{37}$ Liberal Arts Department, School of the Art Institute of Chicago, Chicago, IL, USA 60603\\
$^{38}$ Jet Propulsion Laboratory, California Institute of Technology, Pasadena, CA 91109, USA\\
$^{39}$ Center for Astrophysics $|$ Harvard \& Smithsonian, 60 Garden Street, Cambridge, MA 02138, USA\\
}
\date{Accepted XXX. Received YYY; in original form ZZZ}

\pubyear{2021}

\begin{document}
\label{firstpage}
\pagerange{\pageref{firstpage}--\pageref{lastpage}}
\maketitle

\begin{abstract}
    We search for the signature of cosmological shocks in stacked gas pressure profiles of galaxy clusters using data from the South Pole Telescope (SPT). Specifically, we stack the latest Compton-$y$ maps from the 2500 deg$^2$ SPT-SZ survey on the locations of clusters identified in that same dataset. The sample contains 516 clusters with mean mass $\langle\Mtwohm\rangle = 10^{14.9} \msun$ and redshift $\langle z\rangle = 0.55$. We analyze in parallel a set of zoom-in hydrodynamical simulations from \textsc{The Three Hundred} project. The SPT-SZ data show two features: (i) a pressure deficit at $R/\Rtwohm = 1.08 \pm 0.09$, measured at $3.1\sigma$ significance and not observed in the simulations, and; (ii) a sharp decrease in pressure at $R/\Rtwohm = 4.58 \pm 1.24$ at $2.0\sigma$ significance. The pressure deficit is qualitatively consistent with a shock-induced thermal non-equilibrium between electrons and ions, and the second feature is consistent with accretion shocks seen in previous studies. We split the cluster sample by redshift and mass, and find both features exist in all cases. There are also no significant differences in features along and across the cluster major axis, whose orientation roughly points towards filamentary structure. As a consistency test, we also analyze clusters from the {\it Planck} and Atacama Cosmology Telescope Polarimeter surveys and find quantitatively similar features in the pressure profiles. Finally, we compare the accretion shock radius ($\Rshacc$) with existing measurements of the splashback radius ($\Rsp$) for SPT-SZ and constrain the lower limit of the ratio, $\Rshacc/\Rsp > 2.16 \pm 0.59$.
\end{abstract}


\section{Introduction}

Galaxy clusters are massive structures that contain multiple components, of which dark matter, ionized gas, and galaxies are the dominant ones. The dark matter is essentially collisionless and responsive only to gravity, while the ionized gas responds to hydrodynamical and electromagnetic forces in addition to gravitational ones. Galaxies contain stars --- which are collisionless like dark matter --- in addition to dark matter and multi-phase gas, and thus respond to all three forces mentioned above. The interactions within and between the dark matter and ionized gas, the two components that make up $>99\%$ of the mass in the cluster, determine the cluster's internal structure and energetics \citep[see][for a review]{Kravtsov2012ClusterFormation}.

Clusters are also dynamically young, having formed recently in cosmic history ($z < 2$), and are actively accreting matter from their surroundings, with this accretion happening preferentially along directions of the filamentary large-scale structure. Consequently, each of dark matter, ionized gas, and galaxies can be further deconstructed into two sub-components --- one belonging to the fully-collapsed, bound structure and another to the infalling component that originates in the large-scale structure. Naturally, the study of the galaxy cluster outskirts (for any component) is a key part of both astrophysical and cosmological studies as these outskirts are the transition regime between the two sub-components, and contain an abundance of dynamical information about clusters and their interactions with their environment \citep{Walker2019OutskirtsReview}. 

This work focuses on the gaseous component of the clusters, and in particular on the pressure profiles where sharp, shock-like features can arise from interactions between the gas of the two sub-components. We infer these pressure profiles via the thermal Sunyaev-Zel'dovich (tSZ) signature of clusters \citep{Sunyaev1972SZEffect}, which arises from the inverse compton scattering of Cosmic Microwave Background (CMB) photons off energetic electrons in the hot intracluster medium \citep[see][for reviews]{Carlstrom2002SZReview, Mroczkowski2019SZreview}. While cluster thermodynamics have traditionally been studied using X-ray observations, the tSZ has emerged as the more ideal probe for the cluster outskirts as the signal amplitude depends linearly with density, whereas for X-rays this dependence is quadratic.

The study of shocks is highly relevant to cluster-based studies of both cosmology and astrophysics given that they are a critical mechanism during structure formation for converting gravitational potential energy into thermal energy. Shocks can induce significant deviations in cluster pressure profiles, and can set up thermodynamic non-equilibrium conditions that invalidate common assumptions made in estimating hydrostatic cluster masses; these masses are a relevant quantity for doing cosmology with cluster counts \citep[see][for a review]{Allen2011CosmoClusterReview}, while a clear understanding of hydrostatic equilibrium in clusters is also necessary for certain cluster-based tests of modified gravity \citep{Terukina2014ModGravHSE, Wilcox2015ModGravHSE, Sakstein2016ModGravHSE, Haridasu2021ModGravHSE}. Notably, the process of shock heating generates a thermal non-equilibrium between the electrons and ions, which can alter the expected X-ray and tSZ emissions and will consequently need to be considered in analyses that include these cluster outskirts \citep{Fox1997ElectronNE, Ettori1998ElectronNE, Wong2009ElectronNE, Rudd2009ElectronNE, Akahori2010ElectronNE, Avestruz2015ElectronNE, Vink2015ElectronNE}. 

Shocks can also be sources for accelerating cosmic ray electrons via Diffusive Shock Acceleration \citep{Drury1983DSA, Blandford1987ShocksCRs}. Cosmic ray electrons form a non-thermal tail in the energy distribution of the electron population \citep{Miniati2001NonThermal, Ryu2003CosmologicalShockWaves, Brunetti2014ClusterCRsReview}, and cosmic rays in general alter the total pressure support of the system. Near the cluster core, the pressure from cosmic rays has been observationally constrained to be subdominant to the thermal pressure \citep[\eg][]{Ackermann2014CRpressure} but simulations show it can be more prominent at the outskirts \citep{Pfrommer2007CRsImpactSZandXray, Vazza2012ShocksRadioRelics}. 

The location of shock features also depends closely on the mass accretion rate of the cluster and can potentially serve as an observational proxy for the same \citep{Lau2015GasProfileOutskirts, Shi2016ShockMAR, Zhang2020MergerAcceleratedShocks, Zhang2021SplashShock}. The mass accretion rate has strong theoretical connections to key dark matter halo properties like concentration and formation time \citep{Wechsler2002Concentrations}, and has also been shown to have significant correlations with a broad range of halo properties \citep[\eg][]{Lau2021DMCorrelations, Anbajagane2021BaryonImprints}. However, it has remained difficult to infer observationally.

Accurate measurements of gas profiles at the cluster outskirts --- particularly near and beyond the one-to-two-halo transition regime --- improve the modelling needed in studies of the tSZ auto-correlation \citep[\eg][]{Hill2013tSZPowerSpec, Horowitz2017tSZCosmology, Tanimura2021PlancktSZCosmo} as well as tSZ cross-correlations with galaxy and galaxy cluster positions \citep[\eg][]{Hajian2013PlanckWMAPxROSAT, Vikram2017GalaxyGroupstSZ, Hill2018tSZxGroups, Pandey2019GalaxytSZ, Pandey2020tSZCrossForecast}, with weak lensing shears \citep[\eg][]{Ma2015PlanckxCFHTLens, Hojjati2017PlanckxRCSLenS, Osato2018PlanckxRCSLenS, Osato2020PlanckxHSC, Shirasaki2020XraySZLensing, Gatti2021DESxACT, Pandey2021DESxACT}, or with X-ray luminosity \citep{Shirasaki2020XraySZLensing}; such studies can provide strong and complementary constraints on both cosmological and astrophysical parameters.

Certain shock features form a boundary around the gaseous halo and delineate the cold, pristine gas of the infalling regions from the hot, thermalized gas of the bound structure. This boundary thus marks the radius within which galaxies are first affected thermodynamically by the cluster gas, which consequently impacts the galaxies' evolution \citep[\eg][]{Zinger2016QuenchingShocks} via processes like ram-pressure stripping \citep{Boselli2021RPSReview}. The cosmic rays generated by shocks can also potentially explain the still-unconfirmed origins of radio relics in clusters \citep[\eg][]{Vazza2012ShocksRadioRelics, Hong2014ShockSims, Ha2018ShocksSims} in addition to amplifying seed magnetic fields within clusters \citep[see][for reviews]{Dolag2008NonThermalReview, Donnert2018ClustersMagField}. The magnetic fields also have an inherent non-thermal pressure, and so can impact the total pressure support of a cluster and thus, the hydrostatic cluster mass estimates, just like cosmic rays.

\begin{figure}
    \centering
    \includegraphics[width = 0.85 \columnwidth]{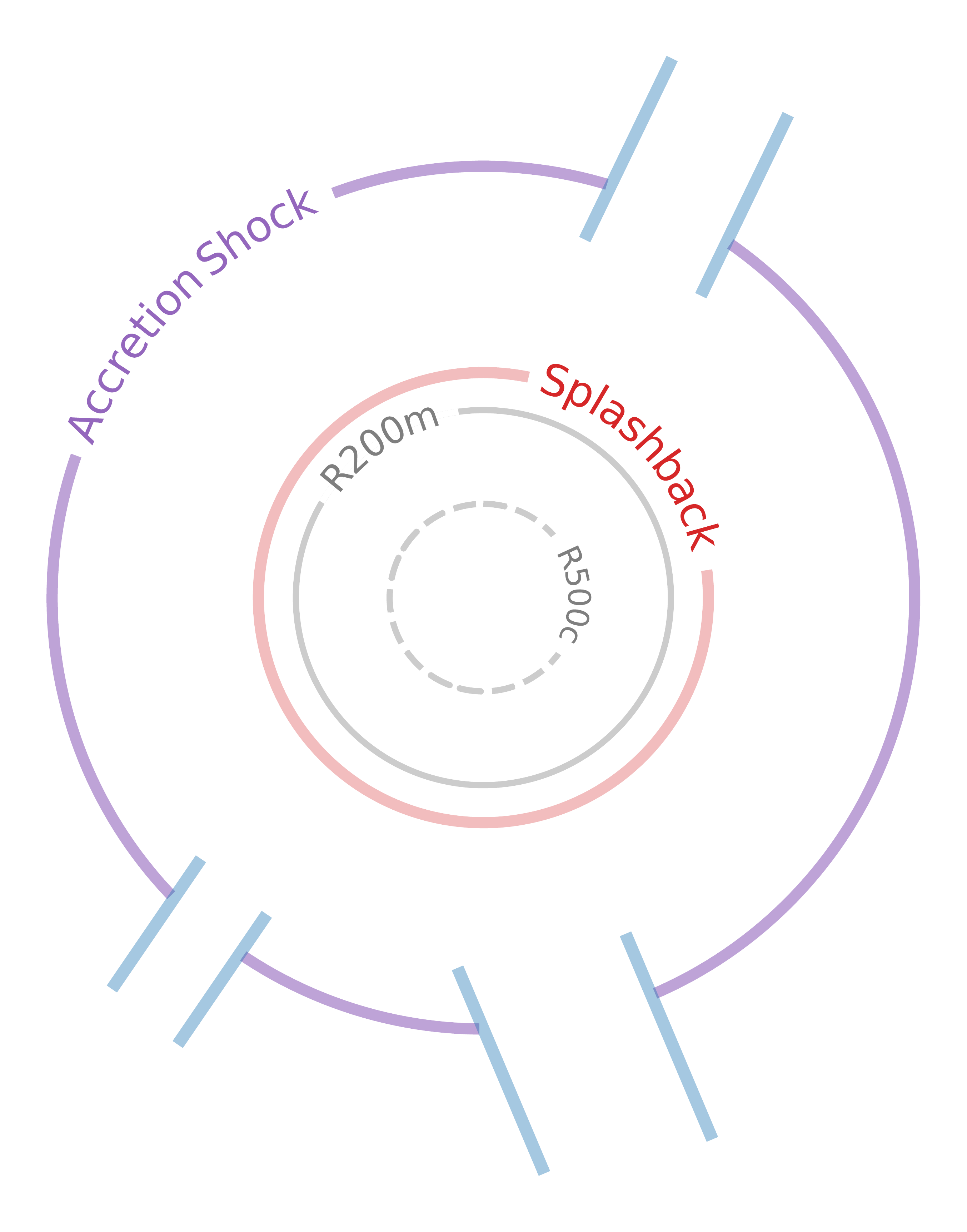}
    \caption{Illustration of the different cluster radii relevant to this work, denoted by colors and labelled by text. Radii are shown on linear axes, and are drawn to relative scales determined for the median SPT cluster mass and redshift --- $\Rfivehc = 0.8 \mpc$, $\Rtwohm = 1.6 \mpc$, $\Rsp = 1.2\Rtwohm = 1.9 \mpc$, and $\Rshacc = 2.3 \Rtwohm = 3.7 \mpc$. The estimate for the ratio $\Rsp/\Rtwohm$ was taken from \citet{Shin2019SplashbackDESxACTxSPT}, while $\Rshacc/\Rtwohm$ was taken from \citet{Baxter2021ShocksSZ}. The blue lines show different filaments connected to the galaxy cluster. The accretion shock is expected to be weak/non-existent along filamentary directions. The splashback radius of the dark matter can be smaller than $\Rtwohm$ depending on the mass accretion rate of the cluster.}
    \label{fig:Shock_Diagram}
\end{figure}

Even more can be learned upon combining the thermodynamic gas structure with the distribution of dark matter and galaxies. One such combination is to compare the shock radii with the splashback radius \citep[\eg][]{Diemer2014Splashback, Adhikari2014Splashback, More2015Splashback, Mansfield2017Splashback, Aung2020SplashShock, Xhakaj2020Splashback, ONeil2021SplashbackTNG, Dacunha2021SplashbackTNG}, which is a physically motivated halo boundary defined by the apocenter in the dark matter phase space of the halo. The existence of the splashback feature has been observationally verified by various analyses \citep{More2016SplashbackSDSS, Baxter2017SplashbackSDSS, Chang2018SplashbackDES, Shin2019SplashbackDESxACTxSPT, Zurcher2019SplashbackPlanckxPanStarrs, Murata2020SplashbackHSC, Adhikari2020SplashbackCosmicClock, Shin2021SplashbackDESxACT}, and has been shown to play a role in galaxy formation physics \citep{Baxter2017SplashbackSDSS,Shin2019SplashbackDESxACTxSPT, Adhikari2020SplashbackCosmicClock, Dacunha2021SplashbackTNG}. The ratio of the shock radius and splashback radius, alongside appropriate theoretical models \citep[\eg][]{Shi2016ShockMAR}, can provide observational constraints on both the adiabatic index of the gas and the mass accretion rate of the cluster \citep[\eg][]{Hurier2019ShocksSZPlanck}. These two features are also sensitive to different types of mass accretion --- the shock radius evolves according to smooth accretion, which does not include accretion of subtructure, whereas the splashback radius depends on the total accretion rate \citep{Zhang2021SplashShock} --- so combining the two could potentially constrain the amount of mass accreted via merging substructures. Figure \ref{fig:Shock_Diagram} shows a diagram of the features discussed above in relation to more commonly used cluster radius definitions. 

Hydrodynamical simulations show that shocks can be generated at different radial locations via different mechanisms, and to zeroth order there are two governing processes: (i) the accretion of gas onto the cluster, i.e. the interaction between the ``bound'' and infalling components, and; (ii) the major and minor mergers with gas clumps, galaxies, and other clusters. The accretion of pristine cold gas --- which has a low sound speed and is primarily found in low-density regions such as cosmic voids --- onto the thermalized, bound gas subcomponent results in a shock of a high mach number and discontinuities in the profiles of many thermodynamic quantities such as temperature, entropy, pressure, and density. This shock is oftentimes referred to as an accretion shock \citep[\eg][]{Lau2015GasProfileOutskirts, Aung2020SplashShock, Baxter2021ShocksSZ} or an external shock \citep{Ryu2003CosmologicalShockWaves}, and has a theoretical foundation that goes back many decades \citep{Bertschinger1985SelfSimilar}.

Closer to the cluster core, the supersonic infall of galaxies and gas clumps in the hot, ionized gas leads to a series of bow shocks with weak mach numbers, which are referred to as internal shocks \citep{Ryu2003CosmologicalShockWaves}. Furthermore, \citet{Zhang2019MergerShock, Zhang2020MergerAcceleratedShocks} found that these bow shocks detach from the infalling substructure, leading to a runaway merger shock that then collides with the accretion shock. This generates a new shock, named the \textit{Merger-accelerated Accretion Shock} or MA-shock, that is both further out and longer lived than the original accretion shock. This is a common process during structure formation, and so most shocks observed in the cluster outskirts ($R \gtrsim \Rtwohm$) are expected to be MA-shocks. While the origin of the MA-shock is rooted in merger events, the radial evolution of the feature --- once it has been generated --- depends only on smooth mass accretion \citep{Zhang2021SplashShock}. Finally, we stress that all of these processes detailed above are complex, highly aspherical, and vary significantly from cluster to cluster.

As was noted before, the current picture of shocks has been studied predominantly using hydrodynamical simulations. Initial studies used non-radiative simulations that modelled gas dynamics but did not include any non-gravitational processes such as gas cooling \citep{Quilis1998ShocksSims, Miniati2000ShocksSims, Ryu2003CosmologicalShockWaves, Skillman2008ShocksSims, Molnar2009ShocksInSZ, Hong2014ShockSims, Hong2015ShocksSims, Schaal2015ShocksIllustrisNR}. More recent studies have included the effects of gas cooling and star formation \citep{Vazza2009ShocksSims,Planelles2013ShocksSims, Lau2015GasProfileOutskirts, Nelson2016ShocksSimsMW, Aung2020SplashShock}, and as well as the effects of feedback from supernovae and active galactic nuclei \citep{Kang2007ShocksSims, Vazza2013FeedbackShocksSims, Vazza2014ShocksSims, Schaal2016HydroShocksIllustris, Baxter2021ShocksSZ, Planelles2021ShocksSims}. Some work has self-consistently modelled the evolution of cosmic-rays alongside galaxy formation \citep{Pfrommer2007CRsImpactSZandXray}, while a handful have also used idealized simulations to explore the propogation of shocks and their dependence on merger events \citep{Pfrommer2006ShocksSims, Ha2018ShocksSims, Zhang2019MergerShock, Zhang2020MergerAcceleratedShocks, Zhang2021SplashShock}. These works use a wide variety of hydrodynamical solvers and astrophysical model prescriptions; see \citet{Vazza2011ShocksCodeComparison} and \citet{Power2020ClusterOutskirtsNIFTY} for comparisons of different implementations.

These works are accompanied by observational studies of shocks that have focused predominantly on small samples --- often containing just one object --- of local, low-redshift clusters \citep{Akamatsu2011ElectronNEAbell, Akahori2012BulletClusterNE, Akamatsu2016ElectronNEXray, Basu2016ALMASZ, DiMascolo2019Shocks, DiMascolo2019ShocksBullet, Hurier2019ShocksSZPlanck, Pratt2021ShocksPlanck, Zhu2021ShockXray}. More general studies of gas thermodynamic profiles, without a specific focus on shocks, have normally not pushed beyond $r \gtrsim \Rfivehc$ \citep[\eg][]{McDonald2014SPTGasProfiles, Ghirardini2017ChandraPressureProfile, Romero2017MUSTANGPressure, Romero2018MultineProfile, Ghirardini2018ShockAbellXMM}, though some do exist \citep{Planck2013PressureProfiles, Sayers2013SZBolocam, Sayers2016BolocamPlanck, Amodeo2021ACTxBOSS, Schaan2021ACTxBOSS}.

With the advances made by many modern CMB experiments, tSZ maps have achieved significantly improved angular resolution and lower noise levels, and this has happened alongside the construction of large catalogs of clusters ($N \sim 10^3 - 10^4$) by multiple different surveys. Together, these improvements enable population-level analyses of the shock features in the thermodynamic gas profiles, and especially of any features in the cluster outskirts ($R \gtrsim \Rtwohm$). We present here such an analysis of $N = 516$ galaxy clusters from the SPT-SZ survey, a 2500 deg$^2$ survey conducted with the South Pole Telescope (SPT). This is a companion work to \citet[henceforth Paper I]{Baxter2021ShocksSZ}, and to our knowledge is the first \textit{observational} population-level study of such features in the cluster outskirts. 

The key goals of this work are to (i) extract the stacked tSZ profiles of clusters and measure deviations from theoretical expectations, such as those deviations induced by shocks and/or by other non-equilibrium processes, (ii) study the dependence, or lack thereof, of such deviations on cluster mass and redshift, and their variation along the cluster major vs. minor axes, and; (iii) compare the shock radii with the splashback radius. Additionally, our focus on the cluster outskirts naturally provides constraints for the one-to-two halo transition regime. Throughout our analysis, we simultaneously analyze \textsc{The Three Hundred} simulations \citep[\textsc{The300,}][]{Cui2018The300} --- which was also employed by the study in Paper I --- both to test our pipeline and to compare simulation predictions with observations. We also compare the SPT-SZ measurements with public data from {\it Planck} and the Atacama Cosmology Telescope Polarimeter (ACTPol) as a consistency test.

This work is organized as follows: We describe our observational and simulation datasets in \S \ref{sec:Data}, including our procedure for generating mock catalogs. In \S \ref{sec:Measurement_Modeling}, we detail both our measurement procedure for obtaining the stacked pressure profiles as well as the formalism of the theoretical model we compare our results to. Our main results are presented in \S \ref{sec:Results}, while comparisons to external data and the splashback radii are in \S \ref{sec:External_data}. We discuss our findings and conclude in \S \ref{sec:Discussion_Conclusions}.

\section{Data} \label{sec:Data}

In this work, we use data from the SPT-SZ survey to set observational constraints, and simulated clusters from \textsc{The300} to test and validate our analysis pipelines. We also compare our SPT-SZ results to those using data from the \textit{Planck} and ACTPol surveys. We describe each of these datasets below, including a description on how we construct a mock catalog from the simulations to match the SPT-SZ data.

The clusters in our samples are labelled by their spherical overdensity mass, $\Mtwohm$, which is defined as,
\begin{equation} \label{eqn:SO_masses}
    M_{\rm \Delta} = \rho_\Delta\frac{4\pi}{3} R_{\rm \Delta}^3,
\end{equation}
with $\rho_\Delta = 200\rho_m(z)$, where $\rho_m(z)$ is the mean matter density of the Universe at a given epoch. The associated radius is denoted as $\Rtwohm$. Features at the cluster outskirts, such as shocks, follow a more self-similar evolution when normalized by this radius definition \citep{Diemer2014Outskirts, Lau2015GasProfileOutskirts}. 

All three of SPT-SZ, ACTPol, and \textit{Planck} infer $\Mfivehc$ from the integrated tSZ emission around each cluster, while the simulated \textsc{The300} catalogs provide $\Mtwohc$ which is computed directly from the particle data. Both $\Mfivehc$ and $\Mtwohc$ are defined by equation \ref{eqn:SO_masses} but with alternative density contrasts of $\rho_\Delta = 500\rho_c(z)$ and $\rho_\Delta = 200\rho_c(z)$, respectively. Here, $\rho_c(z)$ is the critical density of the Universe at a given epoch. In all cases, we convert masses of alternative definitions into $\Mtwohm$ using the concentration-mass relation from \citet{Diemer2019Concentration} and the publicly available routine from the \textsc{COLOSSUS}\footnote{\url{https://bdiemer.bitbucket.io/colossus/}} open-source python package \citep{Diemer2018COLOSSUS}. We find our results are insensitive to assuming other choices for the concentration-mass relation \citep[\eg][]{Child2018ConcentratioMassRelation, Ishiyama2020UchuuConcentration}.

The tSZ amplitude is reported as the dimensionless $y$ parameter,
\begin{equation}\label{eqn:tSZ_y_def}
    y \equiv \frac{k_B\sigma_T}{m_e c^2}\int n_e T_e dl,
\end{equation}
where $k_B$ is the Boltzmann constant, $\sigma_T$ is the Thomson cross-section, $m_e c^2$ is the rest energy of an electron, $n_e$ and $T_e$ are the electron number density and temperature, respectively, and $l$ is the physical line-of-sight distance. Thus $y$ represents the electron pressure integrated along the line-of-sight. For the rest of this work, we use the terms tSZ and $y$ interchangeably; the former is the physical process of interest, but the latter is the actual measurement provided in the maps.

The tSZ effect corresponds to CMB photons scattering off electrons with a thermal (i.e. Maxwellian) energy/momentum distribution. Analogous effects, called the relativistic SZ (rSZ) and non-thermal SZ (ntSZ), correspond to non-Maxwellian energy distributions and may leak into the measured tSZ \citep{Mroczkowski2019SZreview}. In the rSZ effect, the presence of high-temperature electrons ($T_{\rm e} \gtrsim 5\, {\rm keV}$) requires relativistic corrections to the map-making procedure. However, these corrections are at most $5\%$ and are subdominant to the features discussed in this work \citep[see Figure 1]{Erler2018rSZ}. The ntSZ effect can be generated by a cosmic ray electron population, but is a subdominant effect within $\Rtwohc$ of the cluster, where cosmic rays make up $\lesssim 1\%$ of the total pressure \citep{Ackermann2014CRpressure}. Beyond this radius, the cosmic ray energy fraction is not well constrained. For this work, we assume the ntSZ continues to be a subdominant source in the outskirts, but note that the features we discuss here are qualitatively unaffected even if the ntSZ contaminates the tSZ at the $10\%$ level.

\subsection{The South Pole Telescope SZ (SPT-SZ) Survey} \label{sec:SPT_Data}

SPT-SZ is a $2500 \rm \, deg^2$ survey of the southern sky at 95, 150, and 220 GHz, and was conducted using the South Pole Telescope \citep{Carlstrom2011}. The $y$-map used in our analysis was presented in \citet{Bleem2021SPTymap}, has an angular resolution of $1.25^\prime$, and is made using data from both SPT-SZ and the \textit{Planck} 2015 data release; the former provides lower-noise measurements of the small scales, whereas the latter does the same for larger scales (multipole $\ell \lesssim 1000$). The utilized \textit{Planck} data consists of the 100, 143, 217, and 353 GHz maps from the High Frequency Instrument (HFI). The $y$-map is constructed via the commonly used Linear Combination (LC) algorithm \citep[see][for a review]{Delabrouille2009ReviewLCAlgorithm} which is applied to the maps of different frequencies, and here the weights of the linear combination are chosen so as to minimize the total variance in the output map. The weights are also modified to reduce contamination from the cosmic infrared background (CIB); see Section 3.5 in \citet{Bleem2021SPTymap} for more details. In our analysis, the map is further masked to remove point sources as well as the top 5\% of map regions most dominated by galactic dust.

The associated galaxy cluster catalog is derived from the same data used to construct the y-map and contains 516 clusters that were first identified in \citet{Bleem2015ClusterCatalogSPT}, and with updated redshifts and mass estimates provided in \citet{Bocquet2019ClusterCatalogSPT}. We use the latter, updated catalog for our work. Both the map and the cluster catalog are publicly available.\footnote{\url{https://lambda.gsfc.nasa.gov/product/spt/spt_prod_table.cfm}} Our masses come from the \texttt{M500} column and signal-to-noise ratio (SNR) from the \texttt{XI} column.

We also require accurate estimates of the noise in the $y-$map, which then translates into noise in the pressure profiles estimates, when constructing our mock catalog from the simulations. We estimate this for SPT-SZ using the provided \texttt{half1} and \texttt{half2} maps. Each was constructed using half the SPT-SZ and \textit{Planck} data, and with the same LC procedure described above. Taking the difference of these maps nulls any signal, and results in an accurate estimate of the non-astrophysical, instrument-based noise contribution to the $y$-map,
\begin{equation} \label{eqn:Noise_map}
    \texttt{Noise} = \frac{1}{2}(\texttt{half1} - \texttt{half2}).
\end{equation}
We use this map in conjunction with the simulations to assess the impact of noise on the pressure profiles measured for SPT-SZ clusters. We stress that this noise map lacks astrophysical contaminants such as point-sources and dust, meaning the derived noise estimates exclude astrophysical contributions, but such contaminants are also aggressively masked in our analysis. Note also that this noise map is only used to generate the mock catalog from simulations; in particular, the covariance matrix used in analyzing the observational data is built from the full maps, including all unmasked astrophysical sources.

\begin{figure}
    \centering
    \includegraphics[width = \columnwidth]{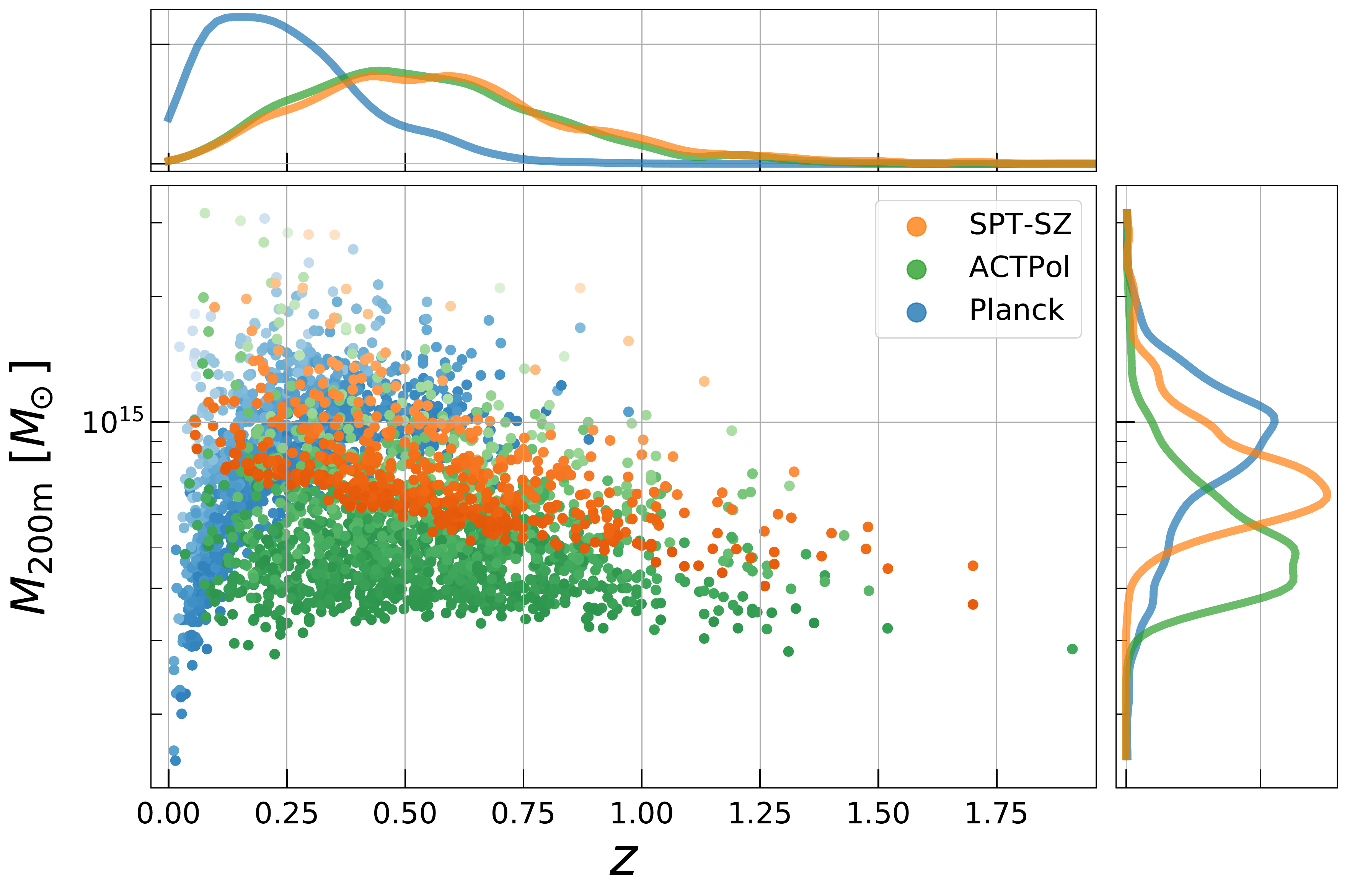}
    \caption{The mass-redshift plane of the cluster samples from SPT-SZ, ACTPol, and \textit{Planck} used in this work. The top and right panels show the 1D distributions for redshift and cluster mass, respectively. The SPT-SZ and ACTPol samples have very similar redshift distributions, with a median of $z \approx 0.55$, while \textit{Planck} is at a lower redshift. Consequently, \textit{Planck} also has more high-mass clusters. The color tones show $\log_{10}\rm SNR$, the signal-to-noise ratio of each cluster detection, with lighter colors indicating a higher SNR. The mean redshift and mass of the different samples can be found in Table \ref{tab:Results}.}
    \label{fig:Characterize_Dataset}
\end{figure}

\subsection{Atacama Cosmology Telescope Polarimeter (ACTPol)} \label{sec:ACT_data}

The ACTPol survey observes at 98 and 150 GHz, and the 2014-2015 observations cover two regions labelled as \texttt{BN} and \texttt{D56}, with an area of $1663 \rm\, deg^2$ and $456 \rm\, deg^2$, respectively. The combined area spans $\approx 2100 \rm\, deg^2$ of the sky. The $y$-map was presented in ACT Data Release 4\footnote{\url{https://lambda.gsfc.nasa.gov/product/act/act_dr4_derived_maps_info.cfm}} \citep[DR4,][]{Madhavacheril2020ACT}, has a resolution of $1.6^\prime$, and makes use of data from both ACTPol and the \textit{Planck} 2015 data release; as was the case with SPT-SZ, the former data inform small-scales and the latter, the large-scales ($\ell \lesssim 1000$). Note that the \textit{Planck} data here consist of eight frequency channels from 30 to 545 GHz, whereas SPT-SZ used four of these channels. Three of the four additional channels used in ACTPol contribute minimally to the final map \citep[see their Figure 5]{Madhavacheril2020ACT}, while the remaining one --- the 70 GHz channel --- has notable contributions below $\ell \lesssim 100$. The map is made separately for each of the two regions using an \textit{Internal} Linear Combination (ILC) algorithm.

In our analysis, the map is further masked to remove point sources and dusty regions. The ACTPol masks are continuous, not binary, and we continue with our aggressive masking by only selecting pixels for which the mask value is 1, meaning the pixel is uncontaminated. Note that the ACTPol maps, unlike all other maps in this analysis, do \textit{not} use the \textsc{Healpy} pixelation scheme and instead work with a new scheme called \textsc{Pixell}.\footnote{\url{https://pixell.readthedocs.io/en/latest/}} Also note that we perform a \textit{combined} analysis of the clusters from each region and do not treat them as separate datasets.

While the public ACTPol $y$-map is part of DR4, the public cluster catalog is from DR5\footnote{\url{https://lambda.gsfc.nasa.gov/product/act/actpol_dr5_szcluster_catalog_get.cfm}} \citep{Hilton2021ACTClusters}, and many DR5 clusters are outside the DR4 y-map footprint.\footnote{We define a cluster to be ``outside the footprint'' if a circle of radius $R = 20\Rtwohm$ around the cluster does not intersect with the footprint. The radius choice reflects the largest scales explored in our analysis, as described in Section \ref{sec:Measurement}.} So, though the full DR5 catalog contains $N_{\rm clusters} \approx 4200$, we are only able to use $N_{\rm clusters} \approx 1400$. As a result, the mass-redshift distribution shown in Figure~\ref{fig:Characterize_Dataset} differs from a similar one shown in \citet[see their Figure 18]{Hilton2021ACTClusters}. Coincidentally, the redshift distribution of the reduced ACTPol cluster sample used in this work is very similar to that of the SPT-SZ sample. For our cluster mass variable, we use the \texttt{M500cCal} column described in \citet[see Table 1]{Hilton2021ACTClusters}, which contains a richness-based, weak-lensing mass calibration factor that brings the ACTPol cluster masses in better agreement with those from SPT-SZ.

\subsection{\textit{Planck}} \label{sec:Planck_data}

The \textit{Planck} satellite mission is a survey of the full-sky and thus, overlaps with both SPT-SZ and ACTPol. The $y$-map, which uses data collected up to 2015, is presented in \citet{Planck2016tSZMaps} and has an angular resolution of $10^\prime$. We use the map constructed using \textsc{Milca} \citep{Hurier2013MILCA} --- a Modified ILC Algorithm --- on the HFI individual frequency maps from the range 100 to 857 GHz. We apply the publicly-available masks to the maps to remove all point sources and an additional 45\% of the sky contaminated by dust.

The cluster catalog was presented in \citet{Planck2016ClusterCatalog}, and has a mass/redshift distribution that significantly differs from both SPT-SZ and ACTPol.
The derived $\Mfivehc$ masses (provided under column \texttt{MSZ}) are also known to be biased low when compared to SPT \citep{Planck2016ClusterCatalog, Battaglia2016ACTxCFHT, Bleem2020SPT-ECS}. By comparing a subset of $94$ clusters that overlap between \textit{Planck} and SPT-SZ, we derive a ratio, $\langle \Mfivehc^{\rm SPT}/\Mfivehc^{\rm Planck} \rangle = 1.28 \pm 0.04$. This ratio is then used to recalibrate the \textit{Planck} masses. We stress that this is only an approximate recalibration, given the mass bias is known to depend on cluster redshift as well \citep[see Section 5.3 in][]{Bleem2020SPT-ECS}. We further discuss the validity and necessity of this procedure in our results (Section \ref{sec:ACT_Planck_Comparison}). We do not explore a more rigorous re-calibration in this work given \textit{Planck} only serves as a comparison point and is therefore not the focus of our analysis.

\subsection{\textsc{The Three Hundred} Project (\textsc{The300})} \label{sec:Simulations}

\textsc{The300} \citep{Cui2018The300} is a set of zoom-in hydrodynamical simulations of 324 massive halos ($\Mfivehc \gtrsim 10^{14.8} \msol$ at $z = 0$). The simulations are run by first identifying the 324 most massive objects in \textsc{MultiDark Planck 2} \citep[MDPL2,][]{Klypin2016MultiDark} --- a $1.5 \gpc$ N-body, dark matter only simulation with purely gravitational evolution --- and then re-simulating those halos with the hydrodynamics code \textsc{Gadget-X} \citep{Beck2016Gadget}, a modified version of \textsc{Gadget3} \citep[last described in][]{Springel2005Gagdet2}, which contains a full prescription of galaxy formation physics such as metal cooling, star formation, kinetic and thermal feedback from supermassive black holes etc. \citep{Rasia2015HydroMethods}. The re-simulated regions are spheres of comoving radii $R = 22 \mpc$ and contain multiple, smaller halos in addition to the most massive halo that was first identified in MDPL2. Halos and subhalos are identified with Amiga's Halo Finder \citep[\textsc{Ahf},][]{Knollmann2009AHF}, which uses an adaptive mesh refinement grid to represent the density field/contours and also has a binding energy criterion to remove unbound material from halos/substructure. The mass computed in AHF follows the $\Mtwohc$ definition, and so we perform the same procedure as in \S \ref{sec:Data} to convert $\Mtwohc \rightarrow \Mtwohm$ for our analysis.

The simulated tSZ/$y$-maps were constructed using the \textsc{PyMSZ} package\footnote{\url{https://github.com/weiguangcui/pymsz}} \citep{Cui2018The300}. At a given redshift, there are 324 maps, each corresponding to an individual re-simulation region, and then we have ten different maps for each region corresponding to the redshifts, $z \in [0.1, 0.2, \ldots, 1.0]$; this redshift range covers $>95\%$ that of the SPT-SZ sample. The maps are also constructed to have the same $1.25^\prime$ angular resolution as SPT-SZ. When comparing observations to simulations, we subsample \textsc{The300} catalogs to replicate the mass and redshift distributions of SPT-SZ. This is done in a hierarchial way --- for each SPT-SZ cluster, we first identify the closest discrete redshift $z \in [0.1, 0.2, \ldots, 1.0]$. Next, we consider all 324 simulated clusters at that redshift and choose the one with an $\Mtwohm$ mass closest to that of the SPT-SZ cluster. Thus, redshift matching takes precedence over mass matching. During such subsampling and comparisons, we limit the SPT-SZ catalog to $z < 1.1$ in order to better match the available $y$-maps from \textsc{The300}, which stop at $z = 1$. This cut excludes only 24 out of the 516 SPT-SZ clusters.

Note that the SPT-SZ sample has 516 \textit{independent} clusters whereas \textsc{The300} follows the \textit{same} set of 324 clusters across different redshifts. So even though we have \textsc{The300} catalogs at multiple redshifts, we only have 324 independent clusters. Thus, during the subsampling step described above, we are forced to select the same clusters from multiple different redshifts, and so our mass- and redshift-matched sample from \textsc{The300} will not be a completely independent set of clusters. This induced correlation will lead to an underestimate of the cluster-to-cluster variance in the pressure profiles; this is only a minor issue as most of the scales we study here are dominated by noise variance instead. 

To ensure we do not study radial scales that extend beyond the $R = 22 \mpc$ spherical volume, we limit our analysis of the simulations to $R < 6\Rtwohm{}$, where only 2 out of the 492 clusters from the mass- and redshift- matched \textsc{The300} sample extend slightly beyond the sphere.

\section{Measurement and Modeling} \label{sec:Measurement_Modeling}

We first describe our procedure for measuring the stacked tSZ profile and other associated quantities in the SPT data, and then the theoretical halo model we compare the measurements with, including how we quantify the significance of any features in the data.

\subsection{Measurement Procedure} \label{sec:Measurement}

We detail below our method for estimating the (i) stacked profiles, (ii) logarithmic derivatives, (iii) bin-to-bin covariance matrix, and; (iv) the feature locations.

\textbf{Estimating stacked profiles:} For each cluster in our catalog, we compute its halo-$y$ correlation, i.e. the average $y$ profile, in 50 logarithmically spaced radial bins that span the range $r \in [0.1, 20]\Rtwohm$. This is done by measuring the average $\MeanY$ in each radial bin and subtracting the mean background value from it, where the latter is estimated by populating the map with random points and computing profiles around them. This entire process is performed using the fast tree-code implementation in the software package \textsc{TreeCorr} \citep{Jarvis2004TreeCorr}. The logarithmic spacing, as opposed to linear spacing, is an apt choice here as the signal (the tSZ emission) drops approximately as a power law with radius, implying a lower SNR per pixel when going from cluster core to the cluster outskirts. Thus, including more pixels per bin towards the outskirts will help increase our SNR per radial bin.

The profiles of the individual clusters are then stacked, with each profile being weighted by the corresponding cluster's detection SNR. Our final results do not change from using alternative weighting choices (see Appendix \ref{sec:Robustness_Tests}). Note that for a given cluster, any radial bin that did not have a single pixel in it --- most commonly the case in the cores of high redshift clusters due to the limited angular resolution --- is ignored during the stacking. In tandem to the stacking procedure, we perform a leave-one-out jackknife resampling, 
\begin{equation} \label{eqn:Jackknife_Mean}
    \MeanY_i(r) = \frac{1}{N - 1}\sum_{j \neq i}^N \,\,y_j(r), 
\end{equation}
where $N$ is the number of measured cluster profiles, $\MeanY_i$ is the mean profile of the sample with cluster $i$ removed, and $y_j$ is the individual profile measurement from cluster $j$. The variance on the mean profile is then given by,
\begin{equation} \label{eqn:Jackknife_Var}
    \sigma^2(\MeanY) = \frac{N - 1}{N}\sum_{j = 1}^N \,\,\bigg(\MeanY_j - \Bar{\MeanY}\bigg)^2,
\end{equation}
where $\Bar{\MeanY}$ is the mean of the distribution of jacknife estimates computed in equation \eqref{eqn:Jackknife_Mean}. Note that equation \eqref{eqn:Jackknife_Var} has an additional factor of $N - 1$ compared to the traditional definition of the variance.

\textbf{Estimating logarithmic derivatives:} Shock features are identified by points of steepest descent in the pressure profiles \citep[\eg Paper I,][]{Aung2020SplashShock}, and this corresponds to measuring minima in the logarithmic derivative. However, derivatives are particularly susceptible to spurious noise-induced features. To alleviate this, the stacked profiles from the jackknife sample are all smoothed by a Gaussian of width $\sigma_{\ln r} = 0.16$, which is $1.5$ times the logarithmic bin width, $\dln r \approx 0.11$. All the profiles we show below have been smoothed by this scale. The smoothing step will induce edge effects at the lower/upper radial limits of $0.1 \Rtwohm$ and $20 \Rtwohm$ given there is no measured profile beyond those bounds. For this reason, we only quote results for the range $0.3 < R/\Rtwohm < 10$. The impact of different smoothing choices, including an essentially un-smoothed case, is discussed in Appendix \ref{sec:Robustness_Tests}, and our results remain the same even when using alternative choices.

We then compute the log-derivative of the smoothed mean profile, ${\rm d}\ln\MeanY/{\rm d}\ln(R/\Rtwohm)$, using a five-point method,

\begin{equation}\label{eqn:logder}
    \frac{df}{dx} = \frac{-f(x + 2h) + 8f(x + h) - 8f(x - h) + f(x - 2h)}{12h},
\end{equation}
where $f$ is an arbitrary function of $x$, and $h = {\rm d}\ln r$ is the spacing between the sampling points. The numerical error in this derivative estimator goes as $\mathcal{O}(h^4)$. The uncertainty on the log-derivative is estimated by computing equation \eqref{eqn:logder} for every stacked profile in the jackknife sample and taking the standard deviation of the resulting distribution. An extra multiplicative factor of $\sqrt{N - 1}$ is then needed to convert the measured uncertainty to the true uncertainty, and this is entirely analogous to the extra $N - 1$ factor in equation \eqref{eqn:Jackknife_Var}.

\textbf{Covariance of the log-derivative:} We also need the bin-to-bin covariance matrix, $\mathcal{C}$, of the log-derivatives when computing a detection significance, as is discussed further below in equation \eqref{eqn:chi2_significance}. This covariance is estimated using the stacked profiles of the jackknife sample,
\begin{equation} \label{eqn:Jackknife_Covar}
    C_{i,j} = \frac{N - 1}{N}\sum_{k = 1}^N \,\,\bigg(f^\prime_{k, i} - \langle f^\prime\rangle_i\bigg)\bigg(f^\prime_{k, j} - \langle f^\prime\rangle_j\bigg),
\end{equation}
where $i$ and $j$ index over the different radial bins, $f^\prime_{k, i}$ is the log-derivative of the $k^{\rm th}$ stacked profile in the $i^{\rm th}$ radial bin, and $\langle f^\prime \rangle_{i}$ is the mean log-derivative in the $i^{\rm th}$ radial bin. For $i = j$, equation \eqref{eqn:Jackknife_Covar} reduces to equation \eqref{eqn:Jackknife_Var}, but for the log-derivatives instead of the profiles.

\textbf{Quantifying feature location:} To determine the location of a given feature --- particularly, of local minima in the log-derivative --- we fit cubic splines to the log-derivative of each stacked profile in the jackknife sample and locate the feature of interest in each. The mean and standard deviation of the resulting distribution provides an estimate for the location of the feature and the associated uncertainty. The $\sqrt{N - 1}$ factor is needed once again to go from the measured uncertainty to the true uncertainty.

Note that our estimates of the locations, $R/\Rtwohm$, do not include the uncertainties in the inferred $\Rtwohm$ of each cluster. For the cluster samples in this work, these uncertainties in $\Rtwohm$ are $<7\%$, which are tolerable given they increase the total uncertainty in the estimated feature location by $<2\%$.

\subsection{Modeling and Detection Quantification} \label{sec:Model_detection_significance}

In addition to comparing our observational results to those from simulations, we also compare the former with commonly used theoretical models, and this will be key in quoting a detection significance for any interesting features. The model we employ here for the halo-$y$ correlation consists of two components: a one-halo term given by the projected version of the pressure profile from \citet{Battaglia2012PressureProfiles}, who calibrated the profiles using hydrodynamical simulations, and a two-halo term  that accounts for contributions from nearby halos as described in \citet{Vikram2017GalaxyGroupstSZ} and later in \citet{Pandey2019GalaxytSZ}. Our two-halo term modelling is based on the linear matter power spectrum and linear halo bias, and assumes higher-order corrections are not required. We validate this assumption below by showing that the linear model accurately describes the profiles of the simulated cluster populations. Our entire modelling procedure is done using the \textsc{Core Cosmology Library (CCL)} open-source python package\footnote{\url{https://github.com/LSSTDESC/CCL}} \citep{Chisari2019CCL}.

We start by computing the 3D, halo-pressure cross-correlation function as the sum of the one-halo and two-halo components,
\begin{equation} \label{eqn:TotHaloDecomposition}
    \xi_{h,p}(r , M, z) = \xi_{h,p}^{\rm one-halo}(r , M, z) + \xi_{h,p}^{\rm two-halo}(r , M, z),
\end{equation}
where $\xi$ are the correlation functions, $r$ is \textit{comoving} distance, and $M$ is the halo mass. We will henceforth denote the combined one-halo and two-halo term as the ``total halo model''. The one-halo term is modelled using the pressure profile from \citet{Battaglia2012PressureProfiles},
\begin{equation} \label{eqn:Battaglia_Prof}
    P(x) = P_{\rm 500c}P_0\bigg(\frac{x}{x_c}\bigg)^\gamma\bigg[1 + \bigg(\frac{x}{x_c}\bigg)^\alpha\bigg]^{-\beta},
\end{equation}
where $P_0$, $x_c$, $\alpha$, $\beta$, and $\gamma$ are the fit parameters calibrated from hydrodynamical simulations, $x = r/\Rfivehc$ is the distance in units of cluster radius, and $P_{\rm 500c}$ is the thermal pressure expectation from self-similar evolution,
\begin{equation} \label{eqn:SelfSimPressureScale}
    P_{\rm 500c} = 500\rho_c(z)\frac{\Omega_b}{\Omega_m}\frac{G\Mfivehc}{2\Rfivehc}.
\end{equation}
Note that while the theory includes the self-similar pressure, $P_{\rm 500c}$, the actual pressure profile normalization is still given by the combination $P_{\rm 500c}P_0$ and accounts for deviations from self-similar evolution via the calibrated parameter $P_0$. The fit parameters for equation \eqref{eqn:Battaglia_Prof} are obtained from the ``Shock Heating'' calibration model of \citet[see Table 1]{Battaglia2012PressureProfiles}, and these parameters have a known, calibrated scaling with both cluster redshift, $z$, and cluster mass, $\Mfivehc$.\footnote{When computing these theoretical predictions for \textsc{The300} clusters, we first convert the provided $\Mtwohc$ masses to $\Mfivehc$ using the same process described in Section \ref{sec:Data} and then use the latter as the input to get the parameters of the theory model.} 

The tSZ emission we study here is connected to the \textit{electron} pressure, $P_e$, whereas the Battaglia profiles are calibrated to the total gas pressure, $P$, so we convert between the two as,
\begin{equation} \label{eqn:Electron_Pressure}
    P_e(r , M, z) = \frac{4 -2Y}{8 - 5Y}P(r , M, z),
\end{equation}
with $Y = 0.24$ being the primordial helium mass fraction. This expression now serves as our one-halo term,
\begin{equation} \label{eqn:OneHalo}
    \xi_{h,p}^{\rm one-halo}(r , M, z) = P_e(r , M, z).
\end{equation}

The two-halo term is more conveniently computed in Fourier space, so we perform all our computations in the same and then inverse Fourier transform in the end to get the correlation function. The two-halo term of the halo-pressure cross-power spectrum, $P^{\rm two-halo}_{h,p}$, is written as,
\begin{equation} \label{eqn:PowerSpectrum}
    \begin{split}
        P^{\rm two-halo}_{h,p}(k , M, z) & = \bigg[b(M, z) \,P_{\rm lin}(k , z) \,\,\times \\
            &\qquad \int_0^\infty \kern-1em dM^\prime \, \frac{dn}{dM^\prime} \, b(M^\prime, z) \,u_p(k,M^\prime, z)\,\bigg],
    \end{split}
\end{equation}
where $M$ is the mass of the halo we are computing the halo-pressure correlation for, $M^\prime$ is the mass of a neighbouring halo contributing to the two-halo term, $P_{\rm lin}(k , z)$ is the linear matter density power spectrum at redshift $z$, $dn/dM^\prime$ is the mass function of neighbouring halos, and $b(M , z)$ and $b(M^\prime , z)$ are the linear bias factors for the target halo and neighboring halos, respectively. The mass function model comes from \citet{Tinker2008HMF} and the linear halo bias model from \citet{Tinker2010HaloBias}. The term $u_p(k,M^\prime, z)$ is the Fourier transform of the pressure profile about the neighboring halo which, under the assumption of spherical symmetry, is computed as,
\begin{equation}\label{eqn:Battalia_Fourier}
    u_p(k,M^\prime, z) = \int_0^\infty dr 4\pi r^2 \frac{\sin(kr)}{kr}P_e(r,M^\prime, z),
\end{equation}
where $P_e$ is the electron pressure profile. The halo-pressure two-point cross-correlation is then simply the inverse Fourier transform of the cross-power spectrum,
\begin{equation} \label{eqn:TwoHalo}
    \xi_{h,p}^{\rm two-halo}(r , M, z) = \int_0^\infty \frac{dk}{2\pi^2}k^2\frac{\sin(kr)}{kr} P_{h,p}^{\rm two-halo}(k , M, z).
\end{equation}
The terms shown in equations \eqref{eqn:OneHalo} and \eqref{eqn:TwoHalo} can be combined according to equation \eqref{eqn:TotHaloDecomposition} to get the total halo model, $\xi_{h,p}$.

We have thus far considered the real-space 3D pressure, whereas the Compton-y parameter is a measure of the \textit{integrated} (or projected) pressure along the line of sight. Thus, the halo-y correlation is given by a projection integral,
\begin{equation} \label{eqn:ProjTotalHalo}
    \xi_{h, y}(r,M, z) = \frac{\sigma_T}{m_e c^2}\int_{-\infty}^{\infty}\frac{d\chi}{1 + z}\xi_{h, p}\bigg(\sqrt{\chi^2 + r^2},M, z\bigg),
\end{equation}
where $\sigma_T$ is the Thomson scattering cross-section, $m_ec^2$ is the rest mass energy of the electron, and $\chi$ is the comoving coordinate along the line-of-sight. 

The tSZ maps we use here have finite angular resolution, which suppresses power on small scales, and so we incorporate this resolution limit into our theory predictions. To do so, we first calculate the angular cross-power spectrum, using the flat sky approximation, as,
\begin{equation} \label{eqn:C_ellConvert}
    C_\ell = \int  \,d\theta\, 2\pi \theta\, J_0(\ell \theta)\, \xi_{h, y}(\theta ,M, z),
\end{equation}
where $J_0$ is the zeroth-order Bessel function. We then multiply $C_\ell$ by the Fourier-space smoothing function for SPT-SZ and then perform an inverse-harmonic transform,
\begin{equation} \label{eqn:TotalHaloSmoothed}
    \xi^{\rm smooth}_{h, y}(\theta , M) = \int \frac{d\ell\ell}{2\pi}J_0(\ell \theta)C_\ell B_\ell,
\end{equation}
with the smoothing function $B_\ell$ given as
\begin{equation}\label{eqn:Beam_Smoothing}
    B_\ell = \exp\bigg[-\frac{1}{2}\ell(\ell + 1)\sigma_{\rm FWHM}^2\bigg],
\end{equation}
where $\sigma_{\rm FWHM} = \theta_{\rm FWHM}/\sqrt{8\ln 2}$, with $\theta_{\rm FWHM} = 1.25^\prime$ being the full-width half-max of the Gaussian filter used to smooth the SPT-SZ maps. The quantity $\theta_{\rm FWHM}$ will take different values when computing theoretical predictions for the ACTPol and \textit{Planck} surveys, as their smoothing scales differ from that of SPT-SZ (see Section \ref{sec:Data}).

To obtain our final theory curve for a given cluster sample, we compute the smoothed total halo model, $\xi^{\rm smooth}_{h, y}$, for each individual cluster in our catalog, and then perform a weighted stack identical to that done on the data, i.e. where the weights are the SNR of the observed clusters. Finally, we quantify the detection significance, which is the significance of a deviation in the measured log-derivatives away from the theoretical model,
\begin{equation} \label{eqn:Detection_significance}
    \epsilon \equiv \frac{1}{\sigma}\bigg(\frac{\dln y^{\rm obs}}{\dln x} - \frac{\dln y^{\rm th}}{\dln x}\bigg),
\end{equation}
where $\sigma$ here is the uncertainty in the log-derivative measurement. The quantity $\epsilon$ is the number of sigma by which the log-derivative in the data differs from that of the theory.

We also measure a standard chi-squared significance for the feature of interest as a whole,
\begin{equation} \label{eqn:chi2_significance}
    \chi^2 = \bigg(\frac{\dln y^{\rm obs}}{\dln x} - \frac{\dln y^{\rm th}}{\dln x}\bigg)^T \mathcal{C}^{-1} \bigg(\frac{\dln y^{\rm obs}}{\dln x} - \frac{\dln y^{\rm th}}{\dln x}\bigg),
\end{equation}
where $\mathcal{C}$ is the covariance matrix of the log-derivative as defined in equation \eqref{eqn:Jackknife_Covar}. We do not use all 50 radial bins for this calculation and instead limit ourselves to 8 bins surrounding the feature of interest. Given our choice of logarithmically space bins with $d\ln r = 0.11$, the total width of 8 bins centered on some location $r_\star$ is approximately $\Delta r \approx r_\star$. Our results do not vary much when using anywhere from 6 to 10 bins in the calculation instead. Once $\chi^2$ is computed, we then quote the total signal-to-noise, $\chi$, of a feature. We note again that our focus on the log-derivatives as our primary detection measure for shocks is informed by recent simulation studies \citep[\eg Paper I,][]{Aung2020SplashShock}.

\section{Shocks in SPT-SZ} \label{sec:Results}

We first present our fiducial results for the SPT-SZ $y-$profiles in Section \ref{sec:Fiducial_Result}, then study the variation in the observed features with (i) cluster mass and redshift in Section \ref{sec:Mass_Redshift_Evol}, and; (ii) cluster major vs. minor axis in Section \ref{sec:Filaments}.

All bands show $68\%$ uncertainties estimated via the jackknife distribution of stacked profiles. As for the detection significance, we show $\epsilon$ in the figures but quote $\chi$ in our discussions in the text, and these quantities are defined in equations \eqref{eqn:Detection_significance} and \eqref{eqn:chi2_significance}, respectively. The latter is our preferred metric as it is the significance of the entire feature across multiple radial bins, while the former is the single-bin significance.

All constraints on feature locations and their corresponding detection significance are provided in Table \ref{tab:Results}. The measured location of any feature is in principle shifted further out from its ``true'' location due to the effects of smoothing in the maps. However, we have estimated the magnitude of this shift by comparing theoretical models with and without smoothing, and find that for both SPT-SZ and ACTPol this shift is negligible in comparison to the uncertainties. The one exception is \textit{Planck}, for which the shift is significant, and we discuss this further in Section \ref{sec:External_data}.

\subsection{Measurements of Pressure Deficit and Accretion Shock} \label{sec:Fiducial_Result}

\begin{figure}
    \centering
    \includegraphics[width = \columnwidth]{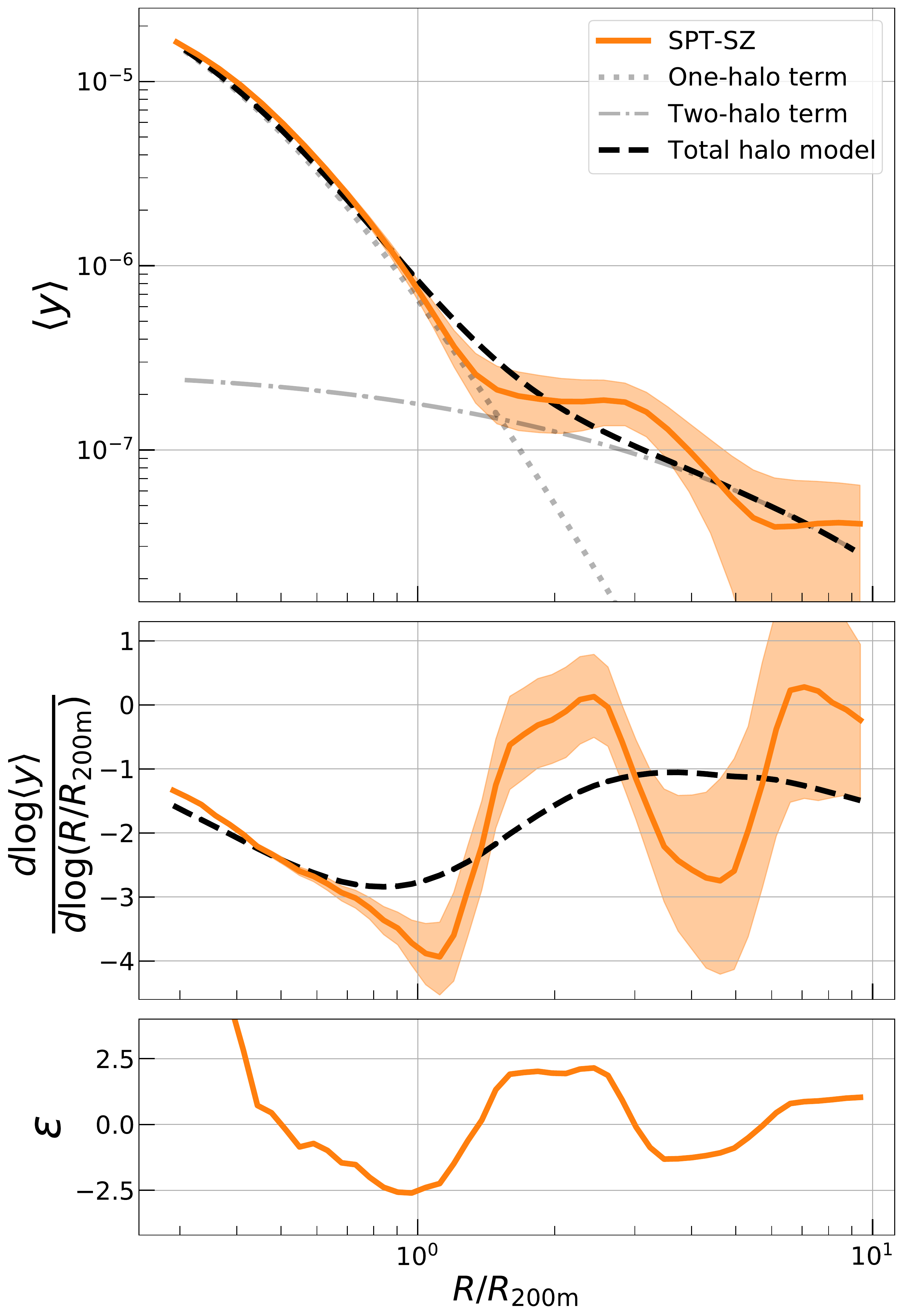}
    \caption{The mean $y$-profile for all $516$ SPT-SZ clusters (top) and the corrsponding log-derivative (middle). The gray dotted (dashed-dotted) line is the one-halo (two-halo) theoretical model described in Section \ref{sec:Model_detection_significance}, while the black dashed line representing the total halo model is a simple sum of the two components. The bottom panel shows the detection significance, described in equation \eqref{eqn:Detection_significance}. The profile shows two features --- the first at $\Rshde/\Rtwohm = 1.08 \pm 0.09$ with $\chi = 3.1\sigma$ significance, and the second at $\Rshacc/\Rtwohm = 4.58 \pm 1.24$ detected at $\chi = 2 \sigma$ significance. The possible physical origins of each feature are discussed in Section \ref{sec:Fiducial_Result}.}
    \label{fig:Detection_of_Shock}
\end{figure}

In general, the stacked $y$-profile of the SPT-SZ clusters follows the trends from the theoretical expectations across both the one-halo and two-halo regime (top panel, Figure \ref{fig:Detection_of_Shock}). There are, however, two clear differences of interest, both found prominently in the log-derivatives. We will refer to these as the pressure deficit and accretion shock, and denote their locations as $\Rshde$ and $\Rshacc$, respectively. Comparisons of these SPT-SZ features with the ACTPol and \textit{Planck} datasets are performed in Section \ref{sec:ACT_Planck_Comparison}.

\subsubsection{Pressure deficit at cluster virial boundary}\label{sec:PressureDeficit}

First, the profile shows a shock-like steepening in the form of a pressure deficit at $\Rshde/\Rtwohm = 1.08 \pm 0.09$ which is not present in the theory, and the corresponding feature in the log-derivatives --- a local minimum --- is measured at $3.1\sigma$ significance. 

This pressure deficit can be a sign of a thermal non-equilibrium between electrons and ions caused by shock heating \citep{Fox1997ElectronNE, Ettori1998ElectronNE, Wong2009ElectronNE, Rudd2009ElectronNE, Akahori2010ElectronNE, Avestruz2015ElectronNE, Vink2015ElectronNE}. Shocks --- which are the primary mechanism for converting kinetic energy to thermal energy during structure formation --- preferentially heat the ions over the electrons given the former are more massive and constitute most of the kinetic energy. This leaves the electrons with a lower temperature than the ions. Normally, Coulomb interactions between the two species re-establish thermal equilibrium, but at the cluster outskirts the particle density is low and the thermal equilibrium time-scale exceeds the Hubble time, as is demonstrated in the above works. Thus, the cluster outskirts will remain in significant non-equilibrium and the temperature difference will consequently impact the tSZ emission, which is sensitive to only the \textit{electron} temperature. \citet[see Figure 2]{Rudd2009ElectronNE} showcase this effect using cluster tSZ profiles from simulations specialized to model the electrion-ion temperature differences, while \citet[see Figure 1]{Avestruz2015ElectronNE} do the same but using the 3D cluster temperature profiles from such specialized simulations. In general, this phenomenon/feature is not resolved in most cosmological hydrodynamical simulations, including all other simulations discussed in this work, given they a-priori assume local thermal equilibrium between electrons and ions.

X-ray observations of local-volume galaxy clusters have shown some indication of thermal non-equilibrium  \citep[\eg][]{Akamatsu2011ElectronNEAbell, Akahori2012BulletClusterNE, Akamatsu2016ElectronNEXray}, though the findings are not yet conclusive and require more data. On the tSZ side, \citet[see Figure 6]{Planck2013PressureProfiles} studied the pressure profiles of 62 clusters from the early \textit{Planck} SZ data and find that their measured profiles are statistically consistent --- given the large uncertainties from the modest sample size --- with many simulations that all assumed local thermal equilibrium.

Other recent works have found deviations in the thermodynamic cluster profiles --- though none explore thermal non-equilibrium as a potential cause. \citet{Hurier2019ShocksSZPlanck} found a sharp decrease in pressure at $R \sim 2\Rfivehc$ for a single cluster in the \textit{Planck} data. \citet{Pratt2021ShocksPlanck} also used \textit{Planck} data and found an \textit{excess} in pressure at $R \sim 2\Rfivehc$ for a set of ten, low-redshift galaxy groups. Finally, \citet{Zhu2021ShockXray} found an excess in the temperature and density profiles of the Perseus cluster at $R \approx \Rtwohc$ using \textit{Suzaku} X-ray data. In all three works, the deviations are found roughly around $R \sim \Rtwohm$, but the exact nature of the deviation --- specifically, whether it is an excess or deficit compared to theory --- varies.

Thermal non-equilibrium is also not the only viable explanation for the observed pressure deficit. The total halo model theory we use was calibrated using a suite of hydrodynamical simulations \citep[see][for more details]{Battaglia2012PressureProfiles}; in principle, any physical process that was not modelled in the simulations could cause a deviation of the observed profiles from the theory. Two notable processes missing in all the simulations we discuss here are magnetic fields and cosmic rays. Both processes can exert a non-thermal pressure, can interact with the ions and electrons, and are particularly relevant in regions containing shocks \citep[see][for a review]{Dolag2008NonThermalReview}. Turbulent gas motions also constitute a significant non-thermal pressure component \citep[\eg][]{Nelson2014NonThermalProfiles, Shi2014NonThermalAnalytic, Shi2015NonThermalSims}, with increasing importance towards cluster outskirts. However, they are unlikely to be the cause of the deviations seen here as such motions, and their resulting non-thermal pressure, are properly resolved in all hydrodynamical simulations.

The top panel of Figure \ref{fig:Detection_of_Shock} shows that the theoretical one-halo term (dotted line) accurately describes the SPT-SZ pressure profiles, \textit{including} the deficit feature. However, upon adding the two-halo term, we find the theory no longer captures the deficit. While one could take this to mean the one-to-two halo transition in the theory is incorrect, this is unlikely to be the case as the profiles in the simulations --- which accurately resolve this transition --- can be modelled precisely using the same theory (bottom row, Figure \ref{fig:Sim_pipeline}). Thus, we do not suspect that this feature arises from a one-to-two halo modelling issue.

\subsubsection{Accretion shocks in cluster outskirts}\label{sec:AccretionShocks}

Next, moving further into the cluster outskirts, the profiles have a nearly constant $\MeanY$ between $\Rtwohm < r < 3\Rtwohm$ before decreasing sharply. This drop is characterized by a local minimum in the log-derivative at $\Rshacc/\Rtwohm = 4.58 \pm 1.24$, and this is measured at $2\sigma$ significance. The radial range of the 8 nearest bins used in the significance computation spans $3 < R/\Rtwohm < 6.4$. While the detection significance is not high, the qualitative behavior is consistent with predictions for an accretion shock from Paper I (see their Figure 5). The location of the minimum is larger, by $\approx 2\sigma$, compared to the findings from previous work on simulations, which locate it at $\approx 2\Rtwohm$ \citep[\eg Paper I;][]{Molnar2009ShocksInSZ, Lau2015GasProfileOutskirts, Aung2020SplashShock}. However, the ``start'' of the feature in the SPT-SZ data --- indicated by where the log-derivative starts decreasing the second time --- is at $R/\Rtwohm = 2.44 \pm 0.25$, whereas the same ``start'' in the simulation predictions is roughly around $R/\Rtwohm \approx 1.9$ \citep[see Paper I;][]{Aung2020SplashShock}. This could imply that the observed feature starts closer to the expected location but is much broader than in past works so the location of the \textit{minimum} in SPT-SZ is further out. For example, the evolution of the shock location over time is known to vary significantly with differences in the mass accretion rate of the cluster \citep[\eg Paper I,][]{Aung2020SplashShock, Zhang2021SplashShock}. The redshift range of the cluster sample we study here is also much wider than that of the past works. 

\subsubsection{Comparisons with simulations}\label{sec:CompareWithSims}

\begin{figure*}
    \centering
    \includegraphics[width = 2\columnwidth]{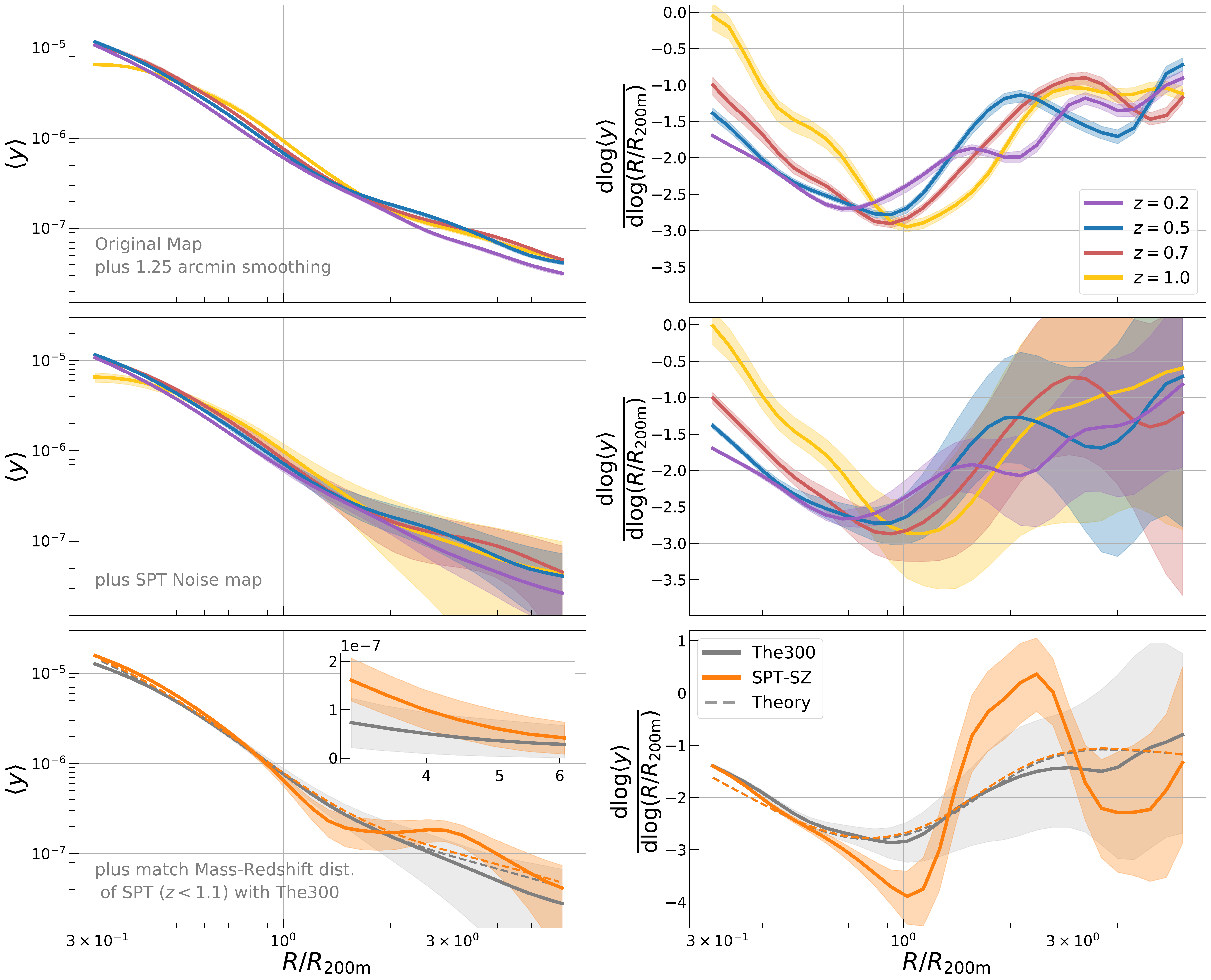}
    \caption{The stacked $\MeanY$ profile (left) and log-derivative (right) for \textsc{The300} simulation across four redshifts (different colors). Top panel show the results from the fiducial maps with $1.25^\prime$ Gaussian smoothing. Middle panels show results from the smoothed plus noise-added maps. The bottom panels use the same maps as the middle panels, but the $\Mtwohm - z$ distribution of the \textsc{The300} cluster sample was matched to the SPT-SZ data. We limit the SPT-SZ cluster catalog to $z < 1.1$ to better match the simulation sample (which only goes up to $z < 1$), and this leave us with 492 clusters. The bands for the simulation results show our estimates for the noise-induced \textit{measurement} uncertainty, and do not represent a theoretical uncertainty. The mass- and redshift-matched sample from \textsc{The300} agrees well with theory, including in the one-to-two halo transition regime. \textsc{The300} profiles do not show any features corresponding to a pressure deficit or accretion shock similar to those seen in the SPT-SZ data. The inset in the bottom left panel shows the profiles for $R/\Rtwohm > 3$, but on linear y-axes, where the absolute uncertainties in the mean profile can be properly compared. See Section \ref{sec:CompareWithSims} for details. The theory curves (dashed lines) in the bottom row are computed as described in Section \ref{sec:Model_detection_significance}.}
    \label{fig:Sim_pipeline}
\end{figure*}

We then perform an explicit comparison between SPT-SZ and \textsc{The300} in Figure \ref{fig:Sim_pipeline}. We first show the mean profiles (left) and log-derivatives (right) and progressively modify the simulated maps with Gaussian smoothing (top row) and SPT-SZ noise (middle row), before comparing results of the mass- and redshift-matched sample from \textsc{The300} with those of the SPT-SZ sample (bottom row). As noted before, \textsc{The300} closely follows the theoretical model and this is not surprising given the pressures profiles of \citet{Battaglia2012PressureProfiles} were themselves calibrated using hydrodynamical simulations.

While we are unable to find an accretion shock feature in the mean profile of \textsc{The300}, previous works have shown that different cosmological simulations, including \textsc{The300}, do resolve accretion shocks \citep[Paper I;][]{Molnar2009ShocksInSZ, Lau2015GasProfileOutskirts, Aung2020SplashShock}. In particular, Paper I used the same \textsc{The300} simulations as we do here and showed that the stacked $y$-profile has a strong accretion shock feature at $R/\Rtwohm \approx 2.3$ at $z = 0.2$, whereas we find none. This is because their study specifically selects relaxed clusters --- using the fraction of total cluster mass contained in substructure as the relaxation criteria --- whereas we perform no such selection as this better represents the SPT-SZ dataset. The inclusion of unrelaxed clusters in the analysis of \textsc{The300} significantly weakens the presence of the accretion shock feature. This could be because relaxation state affects the self-similar scaling of the shock feature with $\Rtwohm$, and deviations can cause the feature to be washed out during stacking. Consequently, the fact that we still see a feature in SPT-SZ indicates that accretion shocks in the data may be stronger than those realized in \textsc{The300}.

Figure \ref{fig:Sim_pipeline} also validates our uncertainty quantification procedure by comparing the uncertainties in the SPT-SZ mean $y-$profile with those of the noise-added simulations, where the noise used in the latter comes from the SPT-SZ half-maps as described in Section \ref{sec:SPT_Data}. Notably, the simulations have a much larger \textit{fractional} (or log) uncertainty than the data, whereas the \textit{absolute} uncertainties of the two agree much better as shown in the inset in the bottom left panel. This is because the fractional uncertainty is amplified by \textsc{The300} mean profile being lower than the SPT-SZ mean profile. 
Note that the cluster-to-cluster variance in the simulations is an underestimate of the true variance as the mass- and redshift-matched sample is not completely independent and contains some correlation. This is due to the same simulated clusters being selected from more than one redshift (see Section \ref{sec:Simulations}).

\subsection{Trends with Mass and Redshift} \label{sec:Mass_Redshift_Evol}

\begin{figure*}
    \centering
    \includegraphics[width = 2\columnwidth]{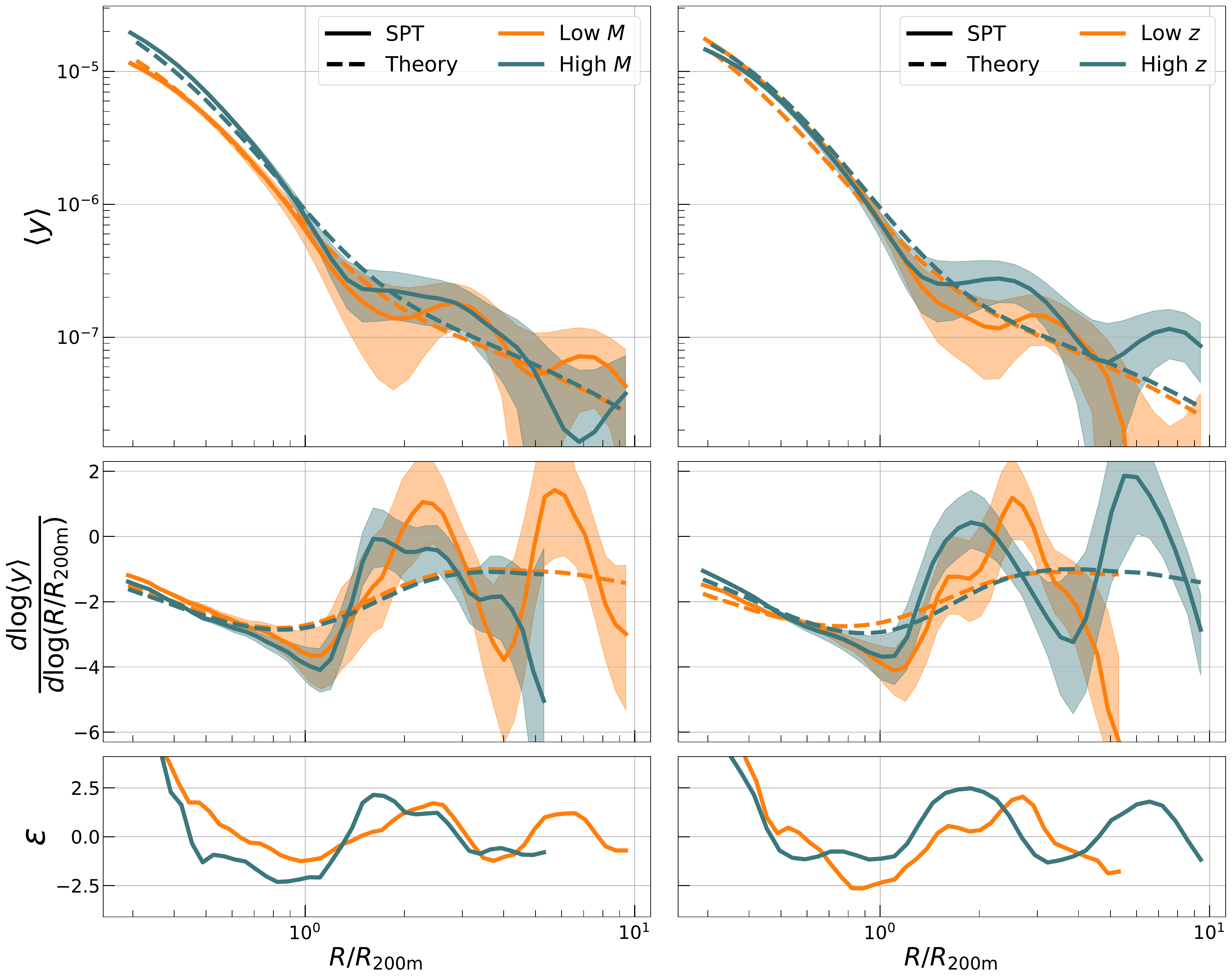}
    \caption{Dependence of the pressure profiles on cluster mass (left) and redshift (right). The panels follow the same style as Figure \ref{fig:Detection_of_Shock}. The subsamples are \textit{not} defined by a simple split in each variable; see Section \ref{sec:Mass_Redshift_Evol} for details. There continues to be a deviation in the log-derivatives at $R \approx \Rtwohm$, and the significance is higher/lower than the fiducial case depending on subsample (see Table \ref{tab:Results}). All subsamples show a plateau phase followed by a sharp drop as is characteristic of an accretion shock feature. We do not observe any statistically significant mass/redshift evolution of either feature.}
    \label{fig:Mass_Redshift_evolution}
\end{figure*}

We next focus on the variation of the two features --- the pressure deficit and the accretion shock --- with cluster mass and redshift. However, the simple approach of splitting the sample according to a median mass or redshift will entangle the effects of mass- and redshift-dependence as the cluster mass evolves significantly over redshift. To properly disentangle these effects, we take a slightly more sophisticated approach which we detail below.

First, to study the temporal evolution, we label each cluster with its peak height
\begin{equation}
    \nu_{\rm 200m} = \frac{\delta_c}{\sigma(\Mtwohm, z)},
\end{equation}
as it is a less redshift-dependent mass definition compared to $\Mtwohm$. Here $\sigma(\Mtwohm, z)$ is the root-mean square fluctuation of the linear matter density field, at redshift $z$, smoothed on the scale $\Rtwohm$, and $\delta_c = 1.686$ is the critical overdensity for collapse. The cluster sample is then rank-ordered according to the peak height labels and then binned into groups of 20 clusters. In each bin, the sample is split into either a high or low redshift group using the median cluster redshift in that bin. The high (low) redshift groups from all peak height bins are combined to form the final high (low) redshift subsample. Thus, each subsample has a similar $\nu_{\rm 200m}$ distribution but different redshift ranges. We use the same measurement procedure as Section \ref{sec:Fiducial_Result}, but now applied to the two subsamples, to get the mean pressure profile and corresponding log-derivative.

The mass evolution is studied using similarly constructed subsamples --- first we rank-order clusters by their redshift, then separate pairs of clusters into high/low mass groups based on $\Mtwohm$, and combine all groups to get the final high/low mass subsamples. In this case, each subsample has a similar redshift distribution but different cluster masses. The resulting mean profile from each subsample is also compared with a theoretical prediction, where the latter now also accounts for the specific redshift- and mass-distribution of the subsample. Table \ref{tab:Results} provides the mean mass and redshift of the subsamples.

The left panels of Figure \ref{fig:Mass_Redshift_evolution} show the results for the mass evolution. The low mass sample has a lower profile normalization and this is the expected behavior from simulations, as is evident from the good agreement between our measurement and the theoretical model for each subsample. The location of the pressure deficit at $\Rshde/\Rtwohm \approx 1.1$ is statistically consistent across all samples, and the significance of the feature varies between $2 \lesssim \chi \lesssim 4$ (see Table \ref{tab:Results}). For the high (low) mass sample we find a $3.8\sigma$ ($1.8\sigma$) significance, indicating the deficit is more strongly observed for higher mass clusters. This is likely just reflecting the fact that higher mass objects have a higher detection SNR. The profiles for both subsamples show a clear plateau in the pressure between $1\Rtwohm < R < 3\Rtwohm$ and a sharp decrease beyond that; this is consistent with the accretion shock feature described in Paper I, as was noted before.

For certain subsamples we do not clearly resolve a second minimum in the log-derivative that would correspond to the accretion shock. The high mass sample in the left panels of Figure \ref{fig:Mass_Redshift_evolution} is one such example. In this case, we only show the log-derivative up to where we have reasonable constraints, i.e. before the error bars blow up due to noise domination, and also only quote a lower limit for the location of the accretion shock. This limit is set by computing the location of the first \textit{maximum} in the log-derivative. Any accretion shock --- identified by the second minimum --- will need to be further out from the location of the first maximum.

The right panels of Figure \ref{fig:Mass_Redshift_evolution} show our results for the redshift evolution. The inner profile of the cluster subsamples are nearly atop one another, whereas the outskirts deviate more but remain statistically consistent. The results seem to suggest that at low redshift the accretion shock feature may be pushed to slightly larger radii, but we do not quantify the significance of this behavior given we do not observe a second minimum in the log-derivative of the low redshift subsample and thus can only place a lower limit on the accretion shock location. The location of the pressure deficit is consistent across subsamples, like before, while the significance varies slightly: the high (low) redshift sample observes the pressure deficit at $2.3\sigma$ ($3.2\sigma$).

A key finding from both the analyses is that the location of pressure deficit is statistically consistent across all subsamples. It is constantly found at $\Rshde/\Rtwohm \approx 1.1$, where the log-derivative is always steeper than in the theory. The characteristic plateauing plus subsequent drop-off of the $y$-profiles, indicative of an accretion shock, also exists in all subsamples.

\subsection{Connection to Filaments} \label{sec:Filaments}

\begin{figure*}
    \centering
    \includegraphics[width = 2\columnwidth]{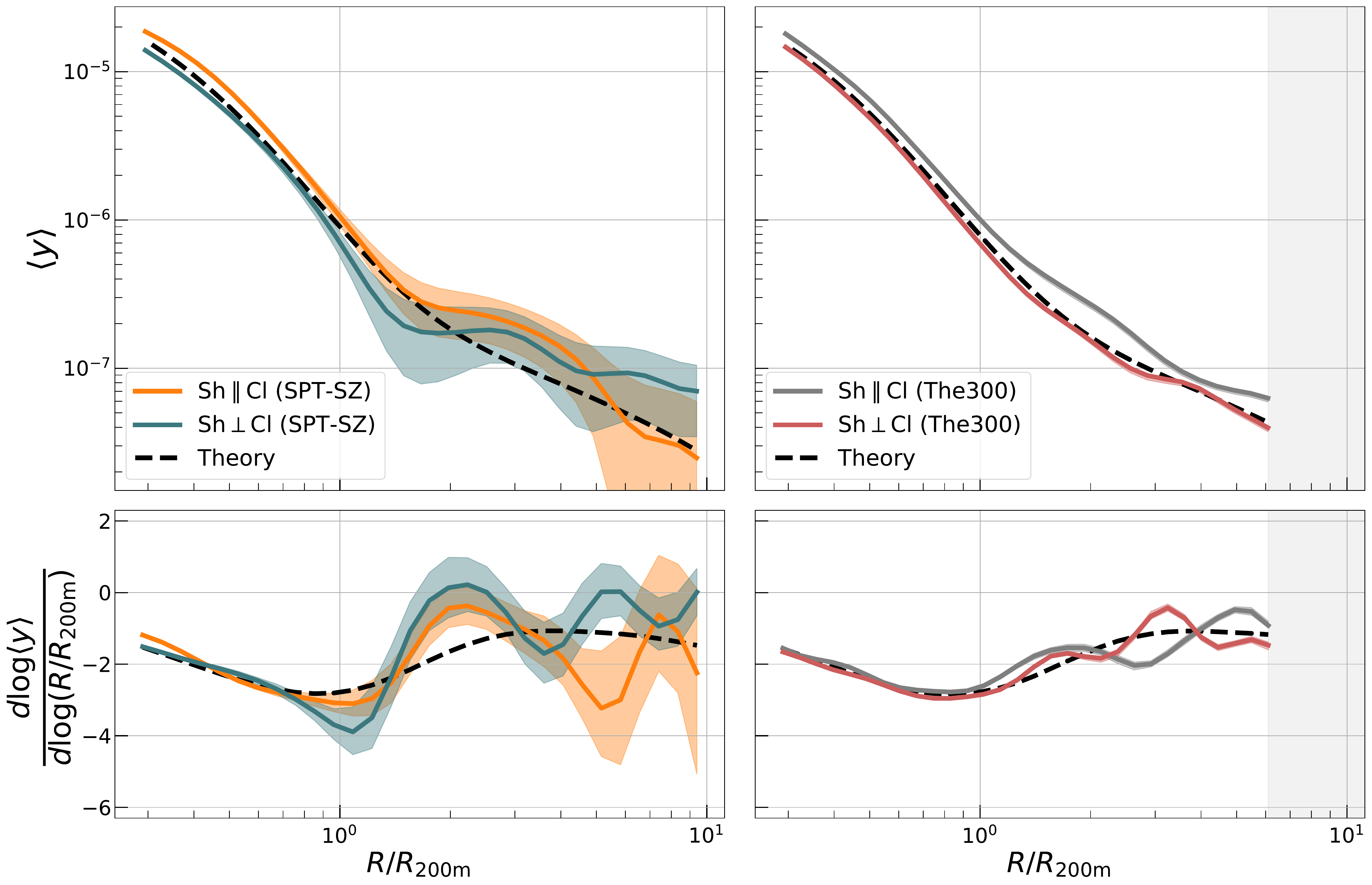}
    \caption{The stacked pressure profiles of SPT-SZ sample (left) and the mass- and redshift-matched \textsc{The300} (right) sample, computed in quadrants that are along/across the cluster major axis (denoted as ``Cl'' in the legend), which we take to be aligned with cosmic filaments. Panels adopt the same format as Figure \ref{fig:Detection_of_Shock}. The profiles are consistent with being stretched along the major axis and squeezed along the minor axis, which leads to a shallower and steeper log-derivative, respectively. The accretion shock along the minor axis is potentially closer to the cluster than the same feature along the major axis, and a similar feature appears to show up in the simulated cluster sample as well. The simulations are cut off at $R < 6\Rtwohm$ due to the limited scales available in the resimulated regions, and these excluded scales are shaded out.}
    \label{fig:Filaments}
\end{figure*}

The morphology of the accretion shock is known to depend on the filamentary structure connected to the galaxy cluster \citep[\eg][]{Molnar2009ShocksInSZ, Lau2015GasProfileOutskirts, Zhang2020MergerAcceleratedShocks, Aung2020SplashShock, Zhang2021SplashShock}. The amount of matter accreted onto the cluster is not isotropic and is preferentially higher along the direction of filaments. However, the gas within filaments is already \textit{preshocked} and is unlikely cold enough to generate an accretion shock \citep[\eg][]{Zinger2016PenetratingGasShocks}. Thus, the accretion shock feature is expected to be anisotropic --- it is a weak-to-nonexistent feature in directions directly towards filaments. However, the accretion shock boundary is also elliptical and has been shown to align with the filamentary large-scale structure \citep[see Figure 1]{Aung2020SplashShock}.

To test this in the data, we assume the cluster major axis  --- as determined by the $y$-map image of the cluster --- is preferentially aligned in the direction with more filamentary structure. So we split this $y$-map image into quadrants, with two quadrants containing the major axis and the other two containing the minor axis. By reorienting and stacking the cluster quadrants appropriately, we compute the mean pressure profile along and across the cluster major axis. Note that the quadrants across the cluster minor axis will still contain filamentary structures, but will contain \textit{fewer} structures than the quadrants along the major axis. The orientation of the cluster in the map is obtained by fitting a 2D Gaussian to the $y$-map using the \textsc{AstroPy} library \citep{Astropy2013, Astropy2018}. We only fit to pixels within $R/\Rtwohm < 0.5$, and find that extending the aperture further leads to systematic biases due to the noise-domination within individual pixels beyond those scales.

The left panels of Figure \ref{fig:Filaments} show the results of such an analysis on the SPT-SZ data. The profile mostly follows our expectations --- along the longer, major axis, it is essentially stretched out, while along the minor axis it is compressed. So at fixed radii, the pressure measured along the major axis would be higher than that along the minor axis. The stretching naturally results in a \textit{shallower} log-derivative, while the squeezing results in a steeper one. We also see features resembling an accretion shock --- i.e. the plateauing followed by a sharp drop off --- but they are not very prominent. The feature along the cluster minor axis is closer in than a similar feature along the cluster major axis, which is consistent with results from \citet{Aung2020SplashShock}. However, note that these are low significance features. We also stress that the theoretical model plotted in Figure \ref{fig:Filaments} does not account for the ellipticity of the halos, and so should not be used to quantify the detection significance of an observed feature. For this reason we quote neither $\epsilon$ or $\chi$ for this analysis.

The right panels of Figure \ref{fig:Filaments} show the results from the mass- and redshift-matched cluster sample from \textsc{The300}. We find similar ``stretching'' and ``compressing'' behaviors in the one-halo term, as expected of clusters that are elliptical, not spherical, in nature. The log-derivatives follow the total halo model theory, though there are some deviations in \textsc{The300} at $r > 2\Rtwohm$ in the form of a second minimum. These could be the accretion shock, but are far too weak to make any strong claims. They are however located around $R/\Rtwohm \approx 2.3$, which is where Paper I found their accretion shock for the relaxed sample of \textsc{The300} clusters at $z = 0.2$. If these two features are indeed accretion shocks, they follow the SPT-SZ data behavior, in that the feature measured along the minor axis is closer to the cluster than that measured along the major axis. The fact that we find some features here, in contrast to finding no clear features in the azimuthally-averaged profiles, also indicates that splitting profiles across cluster major and minor axes may be a better way of searching for accretion shock features.

\section{Comparison with External Data} \label{sec:External_data}

\begin{figure*}
    \centering
    \includegraphics[width = 2\columnwidth]{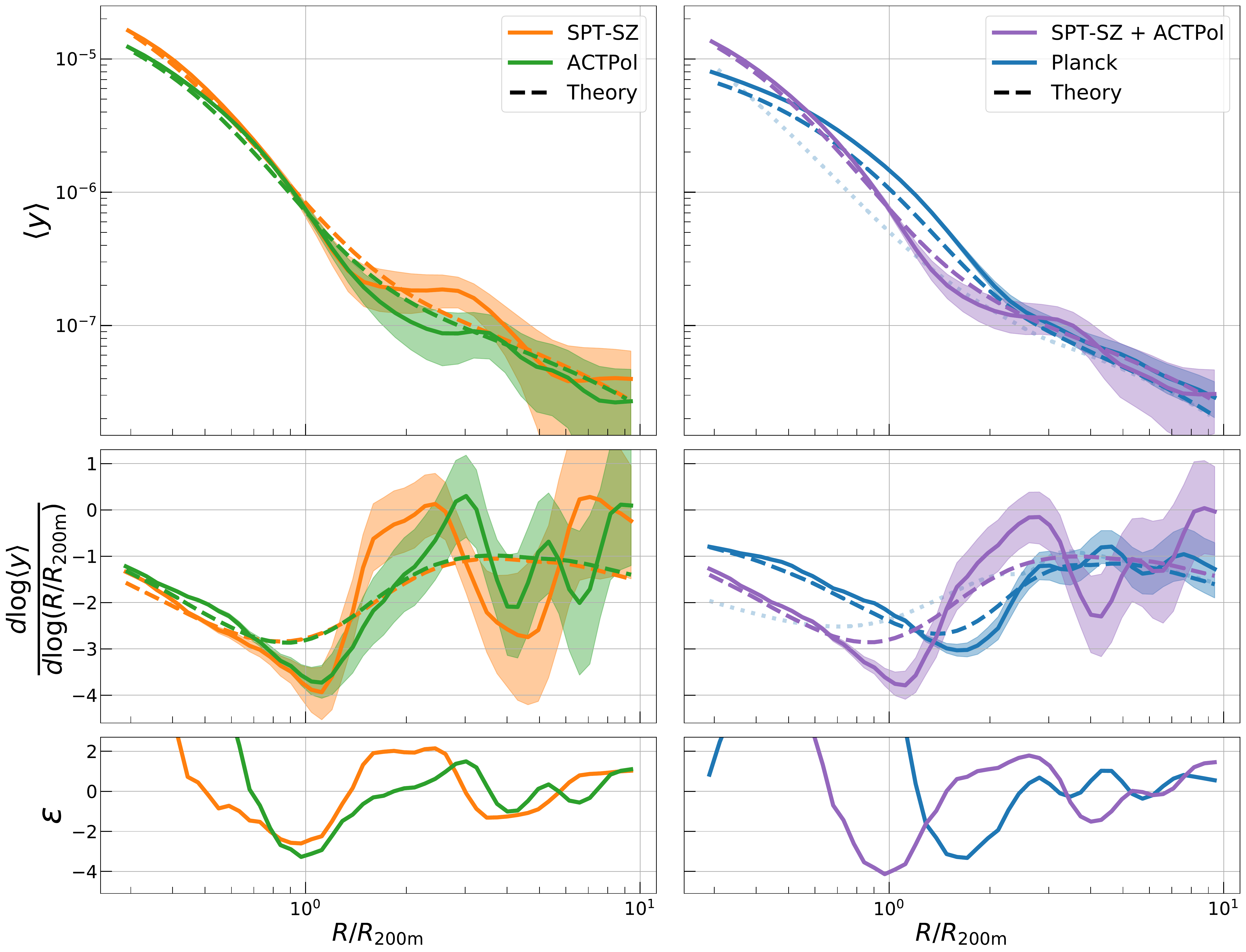}
    \caption{Same format as Figure \ref{fig:Detection_of_Shock}, but for SPT-SZ and ACTPol (left), and for \textit{Planck} and SPT-SZ + ACTPol (right). SPT-SZ and ACTPol show a consistent pressure deficit at $\Rshde/\Rtwohm \approx 1.1$, with $3.1\sigma$ and $3.6\sigma$ significance each. Combining both SPT-SZ and ACTPol data leads to a $4.6\sigma$ significance detection of the pressure deficit, and a $2.7 \sigma$ detection of the accretion shock. \textit{Planck} has a pressure deficit further out at $\Rshde/\Rtwohm = 1.62$, and this is due to its order-of-magnitude larger, $10^\prime$ smoothing scale. The unsmoothed theory prediction for \textit{Planck} is shown in the dotted blue line.}
    \label{fig:Combined_sample}
\end{figure*}

We have thus far focused on the gas pressure profile in the SPT-SZ dataset. In this section, we first repeat our pressure profile analysis using cluster samples and corresponding $y$-maps from two other CMB experiments --- the public ACTPol and \textit{Planck} surveys --- in Section \ref{sec:ACT_Planck_Comparison}. Next, we also compare our SPT-SZ results with existing splashback radius measurements --- made for the same SPT-SZ cluster catalog --- in Section \ref{sec:Splashback_Comparison}. Similar to before, the location and significance of the detected features are summarized in Table \ref{tab:Results}.

\subsection{Consistency with Alternative Datasets} \label{sec:ACT_Planck_Comparison}

Here we repeat the analysis presented in Figure \ref{fig:Detection_of_Shock}, but for three additional cluster samples: (i) ACTPol, and (ii) a combination of SPT-SZ + ACTPol, and; (iii) \textit{Planck}. In each case, the measurements for clusters from a particular survey are made using only the $y$-map released by that particular survey. The descriptions of these maps, and the data used to make them, can be found in Section \ref{sec:Data}.

Comparisons of SPT-SZ with ACTPol and \textit{Planck} provide an important validation step of our findings as the latter two datasets contain a nearly independent set of clusters from SPT-SZ, and their $y$-maps were made using different techniques applied to different data. Note that while the large scales ($\ell \lesssim 1000$) in both SPT-SZ and ACTPol maps are primarily informed by the same \textit{Planck} data, the features we study are localized in real-space, and thus their corresponding harmonic-space features will span a broad range of scales instead of being localized to a range in $\ell$. Additionally, the average angular distance of the pressure deficit feature from the cluster center corresponds to $\ell \sim 2000$, where the SPT-SZ and ACTPol data are more informative than \textit{Planck}. The same for the accretion shock feature corresponds to $\ell \sim 700$.

Figure \ref{fig:Combined_sample} shows the results of this comparison analysis, alongside theoretical predictions constructed in a similar fashion to previous plots. The left column compares SPT-SZ and ACTPol, where we find statistical consistency on both the depth and location of the pressure deficit at $\Rshde/\Rtwohm \approx 1.1$ (see Table \ref{tab:Results} for specific constraints). This provides strong evidence that the feature is of physical origin and also provides a robust consistency test of the feature's radial location. The log-derivatives of ACTPol show some semblance of a second minimum at $4-5\Rtwohm$. Furthermore, the $y$-profiles  from both SPT-SZ and ACTPol showcase a plateauing phase where the pressure is nearly constant over a wide radii range. As previously noted, this feature is consistent with the simulation predictions for an accretion shock as shown in Paper I.

Given that the log-derivatives of the SPT-SZ and ACTPol samples are statistically consistent, we combine the two cluster samples and redo our analysis. The ACTPol and SPT-SZ cluster measurements are still made using only their respective $y-$maps, and the profile data vectors are simply combined to get $N = N_{\rm SPT} + N_{\rm ACT}$ profiles. Note that the SPT-SZ and ACTPol smoothing scales are slightly different ($1.25^\prime$ for SPT-SZ and $1.6^\prime$ for ACTPol) but we have confirmed this difference does not impact our results for the combined sample. The purple lines on the right columns of Figure \ref{fig:Combined_sample} show the results of the combined SPT-SZ + ACTPol sample, where a key finding is the shock-like feature, found at $\Rshde/\Rtwohm = 1.09 \pm 0.06$, is now measured at $4.6 \sigma$ significance. The significance of the accretion shock, which is located at $\Rshacc/\Rtwohm = 4.17 \pm 0.32$, also increases slightly from $2\sigma \rightarrow 2.7 \sigma$. The large reduction of uncertainties in $\Rshacc/\Rtwohm$, from $\pm 1.24$ for SPT-SZ to $\pm 0.32$ for SPT-SZ + ACTPol, is mostly due to the minor change in the shape of the accretion shock feature between the two cases.

Figure \ref{fig:Combined_sample} also shows results from the \textit{Planck} sample, which are significantly different from SPT-SZ and ACTPol primarily due to both the different cluster redshift and mass distributions (see Figure \ref{fig:Characterize_Dataset}), and also the nearly order-of-magnitude larger smoothing scale difference ($\approx 1^{\prime}$ vs. $10^{\prime}$). We still see a pressure deficit in \textit{Planck}, with its location now pushed out to $\Rshde/\Rtwohm = 1.62 \pm 0.14$ due to the smoothing; the dotted blue line in the panel shows the theoretical prediction when no smoothing is included and here the minimum is significantly closer to the cluster core. We estimate the expected shift due to smoothing by measuring the location of the first minimum in the smoothed and unsmoothed theory curves, and find the ratio of locations is $0.5$. This implies the \textit{true}, corrected location of the measured pressure deficit is closer to $\Rshde^{\rm corr}/\Rtwohm = 0.5 \times 1.62 = 0.81$. This pressure deficit is found at $3.0 \sigma$ significance. The uncertainties in both the profile and log-derivative are significantly smaller for \textit{Planck} and this is due to the larger smoothing in the \textit{Planck} $y$-maps.

We do not see any indication of an accretion shock feature for \textit{Planck} in neither the profile nor the log-derivative, and so do not set any lower limits on its location. We suspect this is caused by smoothing, as our tests with the simulations have shown that smoothing at $10^\prime$ significantly suppresses features that were seen at the $1.25^\prime$ smoothing level. Note that there are also minor, nearly-constant vertical offsets between the theory and the observations for \textit{Planck}. This arises because the masses in the \textit{Planck} 2015 cluster catalog are biased low, as was discussed in Section \ref{sec:Planck_data}. We have only approximately corrected for this with our re-calibration step, whereas a full re-analysis of these masses would result in better agreement between observations and theory. For these reasons mentioned above, we do not explore a detailed comparison of \textit{Planck} with SPT-SZ and ACTPol and instead consider the \textit{Planck} results as a more qualitative comparison point.

\subsection{Connecting Splashback and Shock Features} \label{sec:Splashback_Comparison}

\begin{table*}
    \centering
    \begin{tabular}{c|c|c|c|c|c|c|c|c}
        Dataset & $\Rshde/\Rtwohm$ & $\Rshacc/\Rtwohm$ & $\frac{\dln y}{\dln R}(\frac{\Rshde}{\Rtwohm})$ & $\frac{\dln y}{\dln R}(\frac{\Rshacc}{\Rtwohm})$ & $\chi_{\rm sh,\,de}$ & $\chi_{\rm sh,\,acc}$ & $\langle \log_{10}\Mtwohm \rangle[\msol]$ & $\langle z \rangle$\\
        \hline
        \hline
        SPT-SZ & $1.08 \pm 0.09$ & $4.58 \pm 1.24$ & $-3.95 \pm 0.50$ & $-2.77 \pm 1.65$ & $3.1$& $2.0$ & $14.9$ & $0.57$\\
        ACTPol & $1.09 \pm 0.08$ & $> 2.97 \pm 0.21$ & $-3.74 \pm 0.32$ &--- &$3.6$& --- & $14.7$ & $0.54$\\
        Planck & $1.62 \pm 0.14$ & --- & $-3.04 \pm 0.14$ & --- & $3.0$& --- & $15.0$ & $0.25$\\
        \hline
        SPT-SZ + ACTPol & $1.09 \pm 0.06$ & $4.17 \pm 0.32$ & $-3.80 \pm 0.27$ & $-2.40 \pm 0.82$ & $4.6$ & $2.7$ & $14.8$ & $0.55$\\
        SPT-SZ (DES-matched) & $0.93 \pm 0.07$ & $>2.56 \pm 0.41$ & $-3.88 \pm 0.51$ & --- & $3.0$& --- & $14.9$ & $0.49$\\
        \hline
        \hline
        SPT-SZ (High M) & $1.10 \pm 0.10$ & $>2.43 \pm 0.36$ & $-4.11 \pm 0.67$ & --- & $3.8$& --- & $14.9$ & $0.57$\\
        SPT-SZ (Low M) & $1.07 \pm 0.13$ & $4.03 \pm 0.38$ & $-3.69 \pm 0.77$ & $-3.89 \pm 2.48$ & $1.8$& $1.3$ & $14.8$ & $0.57$\\
        \hline
        SPT-SZ (High z) & $1.04 \pm 0.10$ & $3.80 \pm 0.52$ & $-3.77 \pm 0.68$ & $-3.34 \pm 1.96$ & $2.3$& $2.0$ & $14.8$ & $0.68$\\
        SPT-SZ (Low z) & $1.13 \pm 0.12$ & $>2.57 \pm 0.15$ & $-4.15 \pm 0.74$ & --- & $3.2$& --- & $14.7$ & $0.46$\\
        \hline
        SPT-SZ (Major axis) & $1.09 \pm 0.44$ & $5.44 \pm 0.56$ & $-3.11 \pm 0.31$ & $-3.44 \pm 1.65$ & ---& --- & $14.9$ & $0.57$\\
        SPT-SZ (Minor axis) & $1.09 \pm 0.06$ & $3.66 \pm 0.38$ & $-4.53 \pm 0.91$ & $-2.49 \pm 1.15$ & ---& --- & $14.9$ & $0.57$\\
        \hline
    \end{tabular}
    \caption{Table containing the constraints presented in this work. All uncertainties are $\pm1\sigma$ estimates. From left to right the columns show: (i) the sample name, (ii - iii) location of the pressure deficit and accretion shock, (iv - v) the value of the log-derivative at the location of the two features, (vi - vii) detection significance of the two features, extracted using equation \eqref{eqn:chi2_significance}, (viii - ix) the mean log-mass and redshift of the cluster sample. When we do not observe a clear second minimum in the log-derivative, we quote a lower limit for $\Rshacc$ by quantifying the location of the first maximum. We also do not quote the log-derivative or the detection significance in this case. For \textit{Planck}, we cannot set lower limits as we do not observe any accretion shock features. The $\Rshde$ estimate for \textit{Planck} is also strongly impacted by smoothing, making it significantly different from SPT-SZ and ACTPol, and a smoothing-corrected value is provided in the text. We also do not quantify the significance of features in the major/minor axis analyses of SPT-SZ as we do not have an appropriate theory model to use as a baseline.}
    \label{tab:Results}
\end{table*}

\begin{figure}
    \centering
    \includegraphics[width = \columnwidth]{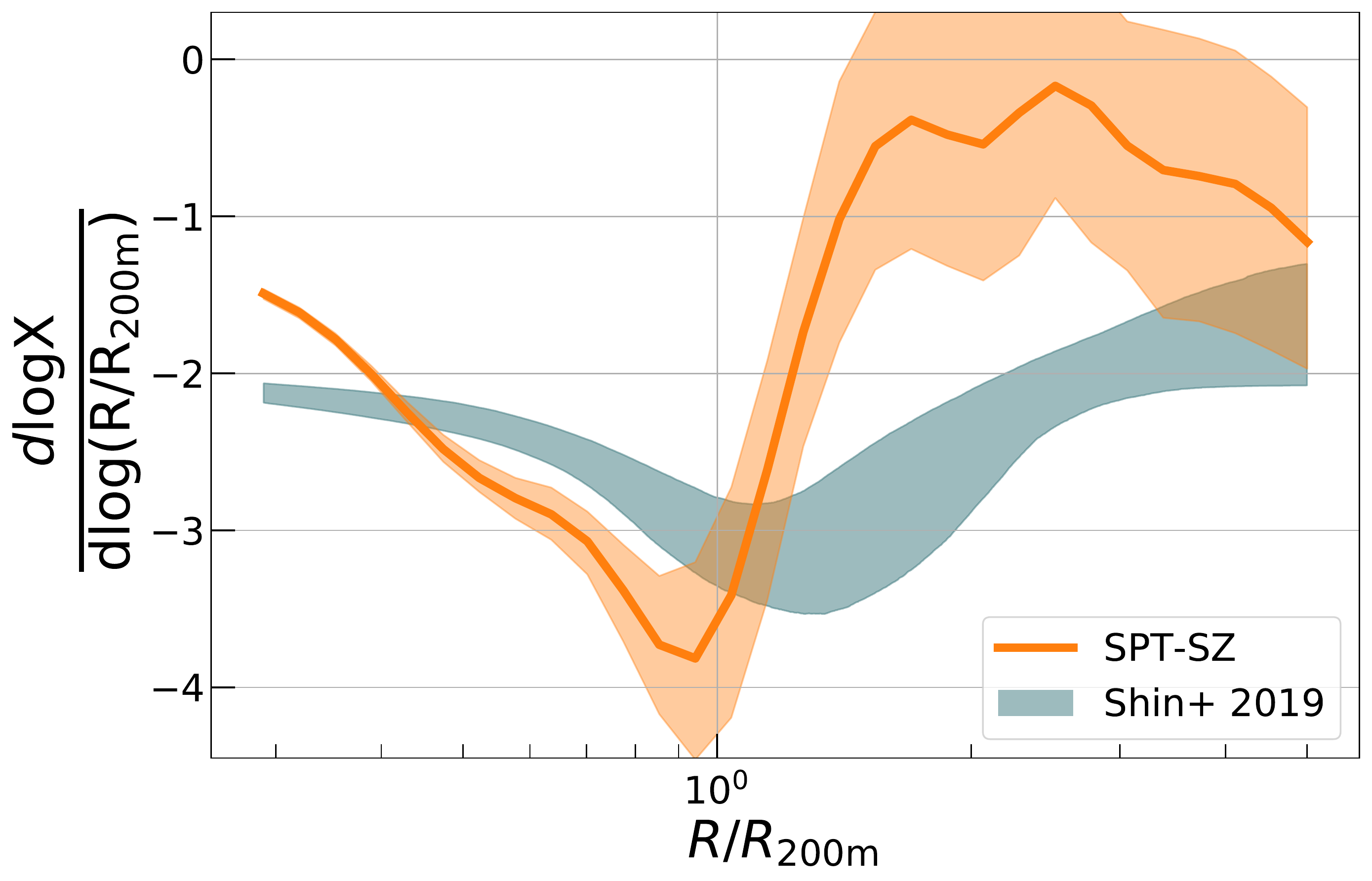}
    \caption{The log-derivative of the mean pressure profile alongside that of the mean galaxy number density profile from \citet{Shin2019SplashbackDESxACTxSPT}. Both measurements are made on the same subset of SPT-SZ clusters, which is different from the data used for Figure \ref{fig:Detection_of_Shock}. See Section \ref{sec:Splashback_Comparison} for details on sample selection. The two profiles have different shapes at all radii, evidenced by the different log-derivatives. The pressure deficit is located inside the splashback radius, but the locations differ by just over $1\sigma$. The accretion shock was not measurable due to increased noise levels from using fewer clusters in this analysis. The location of the pressure profile features are shown in Table \ref{tab:Results}, and a lower limit for the ratio of accretion shock radius and splashback radius is given in Equation \eqref{eqn:SplashShockRatio}.}
    \label{fig:Splashback_comparison}
\end{figure}

Simple, self-similar evolution models of structure formation predict that the locations of the accretion shock and splashback feature coincide for a gas adiabatic index $\gamma = 5/3$ \citep{Bertschinger1985SelfSimilar}. Assuming different gas adiabatic indices in the range $\gamma \in [4/3, 3]$, and also allowing the mass accretion rate to vary, can lead to nearly factor of 2 differences in the location of the accretion shock in self-similar solutions \citep{Shi2016ShockMAR}. However, more complex, 3D hydrodynamical simulations have all found that the accretion shock feature is consistently located well beyond the splashback radius \citep[Paper I;][]{Lau2015GasProfileOutskirts, Aung2020SplashShock, Zhang2020MergerAcceleratedShocks, Zhang2021SplashShock}, and this is related to the MA-shocks we discussed earlier \citep{Zhang2020MergerAcceleratedShocks}.

Here, we compare features in the gas pressure profiles presented in this work with the dark matter splashback feature previously measured for the SPT-SZ sample by \citet[S19]{Shin2019SplashbackDESxACTxSPT}. In S19, the splashback feature is measured by correlating SPT-SZ galaxy clusters with a galaxy sample from the Dark Energy Survey \citep[DES,][]{Flaugher2005} and constructing galaxy density profiles that serve as a proxy for the shape of the dark matter radial density distribution. The minimum in the log-derivative of that profile provides the location of the splashback features; in this sense the splashback and shock measurements are highly analogous. 

S19 were unable to use all 516 SPT-SZ clusters as they required clusters to be in both the SPT-SZ and DES footprints, and to be in the redshift range of the DES galaxy sample, $0.25 < z < 0.7$. In order to perform a fair comparison between works, we use the same selections as S19 on our cluster sample and redo the stacking and log-derivative measurements for this new subsample. Specifically, we only use clusters that are in the redshift range $0.25 < z < 0.7$ and that are contained in the DES footprint. This selection lowers our sample from $N = 516 \rightarrow 256$.

Figure \ref{fig:Splashback_comparison} compares log-derivatives of the pressure profiles and of the galaxy number density profiles of SPT-SZ clusters. We change our upper radial limit to $R/\Rtwohm = 10 \rightarrow 5$ as the factor of 2 reduction in sample size, due to selection cuts, makes measurements of the mean pressure profile beyond $R/\Rtwohm = 5$ noise-dominated. The log-derivatives, which quantify the shape of the profile, are broadly similar within $R < \Rtwohm$, and deviate significantly at $R > \Rtwohm$. This is consistent with inner profiles being determined to zeroth order by simple gravitational evolution, while at the outskirts the presence of shocks causes significant differences in the behavior of infalling dark matter and gas. The pressure deficit is now found at $\Rshde/\Rtwohm = 0.93 \pm 0.07$ and is statistically consistent with the location of the splashback feature, which is at $\Rsp/\Rtwohm = 1.22 \pm 0.25$ as quoted in S19. While $\Rshde$ in the fiducial sample is further out than the DES-matched sample described here, the difference is a little over $1\sigma$. Moreover, the DES-matched sample has a significantly different redshift distribution --- it excludes 148 high-redshift clusters ($z > 0.7$) and 61 low-redshift ones ($z < 0.25$) --- which could cause the minor difference in $\Rshde$.

We also do not find a visible second minimum in the log-derivatives that could indicate the presence of an accretion shock feature. However, we can still set a lower limit on the radial location of such a feature by quantifying the location of the first maximum, like we did in Section \ref{sec:Mass_Redshift_Evol}. We find $\Rshacc/\Rtwohm > 2.56 \pm 0.41$ for this lower limit. Using this value, we can also set a lower limit on the ratio between the accretion shock radius and the splashback radius, which is 
\begin{equation}\label{eqn:SplashShockRatio}
    \Rshacc/\Rsp > 2.12 \pm 0.59.
\end{equation}

This ratio is statistically consistent with expectations from simulations, which have been in the range $1.5 \lesssim \Rshacc/R_{\rm sp} \lesssim 2.5$ \citep[Paper I;][]{Aung2020SplashShock, Zhang2021SplashShock}. If we assume the true location of the accretion shock in the subsample is roughly the same as its location in the fiducial sample (see Section \ref{sec:AccretionShocks} or Table \ref{sec:Results}), then the ratio is closer to $\Rshacc/\Rsp \approx 4$.

\section{Summary and Discussion} \label{sec:Discussion_Conclusions}

The outskirts of galaxy clusters are where bound material within the cluster interacts with infalling material that is accreted from the filamentary large-scale structure. Naturally, these outskirts are dynamically active and contain many instances of shocks. The study of such shocks is of particular interest given the impact of shocks on a wide array of cluster science; from their relevance to cosmology, via their connection to cluster hydrostatic mass estimates, cluster dynamical states, and modelling of tSZ auto/cross-correlations, to their relevance for astrophysics, via their influence on galaxy formation, radio relics, magnetic fields, and cosmic ray production. These cluster outskirts have only recently become observable due to the availability of large samples of clusters and of $y-$maps with improved angular resolution and noise levels; both are necessary advances for making meaningful measurements in the noise-dominated regimes. In this work, we use 516 galaxy clusters from the SPT-SZ survey, and search for features in the stacked gas pressure profiles between $0.3 < R/\Rtwohm < 10$. Our key quantitative results are all provided in Table \ref{tab:Results}, and our main findings are summarized below:

\begin{itemize}
    \item There is a pressure deficit at $\Rshde/\Rtwohm = 1.08 \pm 0.09$ detected at $3.1\sigma$ (Figure \ref{fig:Detection_of_Shock}), and is consequently missing in both simulations and analytic predictions (Figure \ref{fig:Sim_pipeline}). Past works show that such a feature can be expected from a shock-driven thermal non-equilibrium between electrons and ions, and also suggest that the lack of a similar feature in the simulation-calibrated theory and \textsc{The300} simulations is due to the simulations' assumption that electrons and ions are in thermal equilibrium at all times.\vspace{0.5em}
    
    \item The mean pressure profiles undergo a plateau phase where the pressure is constant between $1 \lesssim R/\Rtwohm \lesssim 3$ and then drops sharply. The corresponding minimum in the log-derivative is located at $\Rshacc/\Rtwohm = 4.58 \pm 1.24$, and measured at a low ($2\sigma$) significance, but the qualitative features are consistent with predictions of an accretion shock from Paper I.\vspace{0.5em}
    
    \item Both the deficit and the plateau features are present when splitting the sample based on mass or redshift, and do not show any statistically significant evolution with either quantity. The pressure deficit is found both along and across the cluster major axis, and does not show any significant differences either. There is a mild indication that the boundary shell formed by the accretion shock is elliptical and points in the same direction as the cluster major axis (and thus, in the direction of filamentary structure).\vspace{0.5em}
    
    \item Comparisons with similarly measured pressure profiles from ACTPol and \textit{Planck} show pressure deficits that are quantitatively consistent with the deficit seen in SPT-SZ. The ACTPol pressure deficit in particular has the same location and depth as the SPT-SZ feature. ACTPol also has qualitative features indicative of an accretion shock, but we see no clear second minimum in the log-derivatives.\vspace{0.5em}
    
    \item Comparing the pressure profiles with the splashback radius estimate of \citet{Shin2019SplashbackDESxACTxSPT}, we find the location of the pressure deficit is statistically consistent with the splashback radius, $\Rsp$. The additional selections needed for this comparison deteriorate the SNR of our measurements, and so we are unable to resolve the accretion shock. However, we place a lower limit on the accretion shock location, which translates to a lower limit on the ratio of shock radius to splashback radius. We find $\Rshacc/R_{\rm sp} > 2.12 \pm 0.59$, which is consistent with previous simulation studies.
    
\end{itemize}

Our work shows that the pressure deficit feature can already be measured with strong significance using existing data. Future studies can probe its existence in other SZ datasets to perform further consistency checks at higher precision and/or, more interestingly, probe the physical origin of the deficit as well. In this work, we have discussed how the deficit could arise from a shock-induced thermal non-equilibrium between ions and electrons. However, other physical processes may also resolve this difference. More detailed analysis of this feature --- on both observational and simulation fronts --- will be necessary to better understand the thermodynamic structure of the cluster outskirts and update our models/assumptions appropriately.

Measurements of the accretion shock feature, on the other hand, are currently still dominated by the instrument noise. We have partially worked around this issue by stacking $N \sim 10^3$ clusters, which significantly lowers the noise amplitude, but this still results in only a weak measurement of the feature. Ongoing and future SZ surveys --- like SPT-3G \citep{Benson2014SPT3G}, Advanced ACT \citep{Henderson2016AdvAct}, Simons Observatory \citep{SimonsObs2019}, and CMB-S4 \citep{CMBS42019} --- will both have a higher sensitivity \textit{and} a larger sample of clusters. Both factors will greatly improve the ability to make precise measurements of the stacked mean pressure profile, leading to more precise measurements of the accretion shock. It will also make it possible to better study the dependence of the accretion shock on redshift, mass, and orientation. Studying the dependence on orientation will also provide insight into how shocks respond to the dynamics of mass accretion, which differs significantly between directions toward filaments and toward voids.

Another consideration --- particularly regarding the accretion shock --- is the relaxation state of the cluster. Studies of shocks will benefit from preferentially selecting systems where the shock feature evolves more self-similarly with cluster radii, i.e. relaxed systems. A potential avenue is to make SZ measurements around X-ray selected clusters, given the latter are preferentially more relaxed. Such a technique is viable with the current eROSITA All-sky X-ray mission \citep{Merloni2012eROSITAScienceBook} which will detect $N \approx 10^{5}$ galaxy groups and clusters, and of a median redshift, $z = 0.35$ \citep{Pillepich2012eRositaForecast}. Selecting relaxed clusters could also be done using the SZ map morphology alone \citep[\eg][]{Capalbo2021ZernikePoly}. Any such selection would allow for an observational measurement that more closely mimics the study of Paper I and improves the strength of the observed accretion shock feature in stacked profiles.

\section*{Acknowledgements}

We thank Tara DaCunha and Abigail Lee for their initial exploratory analyses of tSZ profiles in the Planck dataset. We thank Greg Bryan, Damiano Caprioli, Mark Devlin, Andrey Kravtsov, Congyao Zhang, and Irina Zhuravleva for helpful discussions on gas physics in clusters, Shivam Pandey for discussions on the theoretical modelling of the tSZ profiles, Tae-Hyeon Shin for providing us with the SPT-SZ log-derivative curves from S19, and Maya Mallaby-Kay for kindly helping us navigate the \textsc{Pixell} software library. We also thank the anonymous referee for raising useful points that added to the discussion presented here.  Finally, we are grateful to the SPT-SZ, ACTPol, and \textit{Planck} collaborations for making their data products publicly available to the community. This project strongly benefited from the free-flow of information across surveys and collaborations. We additionally thank Colin Hill and Mathew Madhavacheril for their generous support in including the ACT data in this study.

DA is supported by the National Science Foundation Graduate Research Fellowship under Grant No. DGE 1746045. CC is supported by the Henry Luce Foundation and DOE grant DE-SC0021949. BJ is supported in part by NASA ATP Grant No. NNH17ZDA001N and DOE Grant No. DE-SC0007901. WC is supported by the European Research Council under grant number 670193 and by the STFC AGP Grant ST/V000594/1. He further acknowledges the science research grants from the China Manned Space Project with NO. CMS-CSST-2021-A01 and CMS-CSST-2021-B01. LDM is supported by the ERC-StG `ClustersXCosmo' grant agreement 716762. CR acknowledges support from the Australian Research Council Discovery Projects scheme (DP200101068). AS is supported by the ERC-StG ‘ClustersXCosmo’ grant agreement 716762, by the FARE-MIUR grant 'ClustersXEuclid' R165SBKTMA, and by INFN InDark Grant. The South Pole Telescope program is supported by the National Science Foundation (NSF) through award OPP-1852617

All analysis in this work was enabled greatly by the following software: \textsc{Pandas} \citep{Mckinney2011pandas}, \textsc{NumPy} \citep{vanderWalt2011Numpy}, \textsc{SciPy} \citep{Virtanen2020Scipy}, and \textsc{Matplotlib} \citep{Hunter2007Matplotlib}. We have also used
the Astrophysics Data Service (\href{https://ui.adsabs.harvard.edu/}{ADS}) and \href{https://arxiv.org/}{\texttt{arXiv}} preprint repository extensively during this project and the writing of the paper.

\section*{Data Availability}

All data used in our analyses of SPT-SZ, ACTPol, and \textit{Planck} are publicly available at the repositories linked to in this paper. Data products from \textsc{The300} simulations are not hosted on a public repository, but the interested reader is encouraged to reach out to \textsc{The300} Collaboration for data access.

The code used to generate the theoretical tSZ profile of a halo, including both one-halo and two-halo contributions, is made available at \url{https://github.com/DhayaaAnbajagane/tSZ_Profiles}. This repository also contains the profiles and log-derivatives shown in the figures of this work.

\bibliographystyle{mnras}
\bibliography{References}

\appendix

\section{Tests of Robustness} \label{sec:Robustness_Tests}

\begin{figure*}
    \centering
    \includegraphics[width = 2\columnwidth]{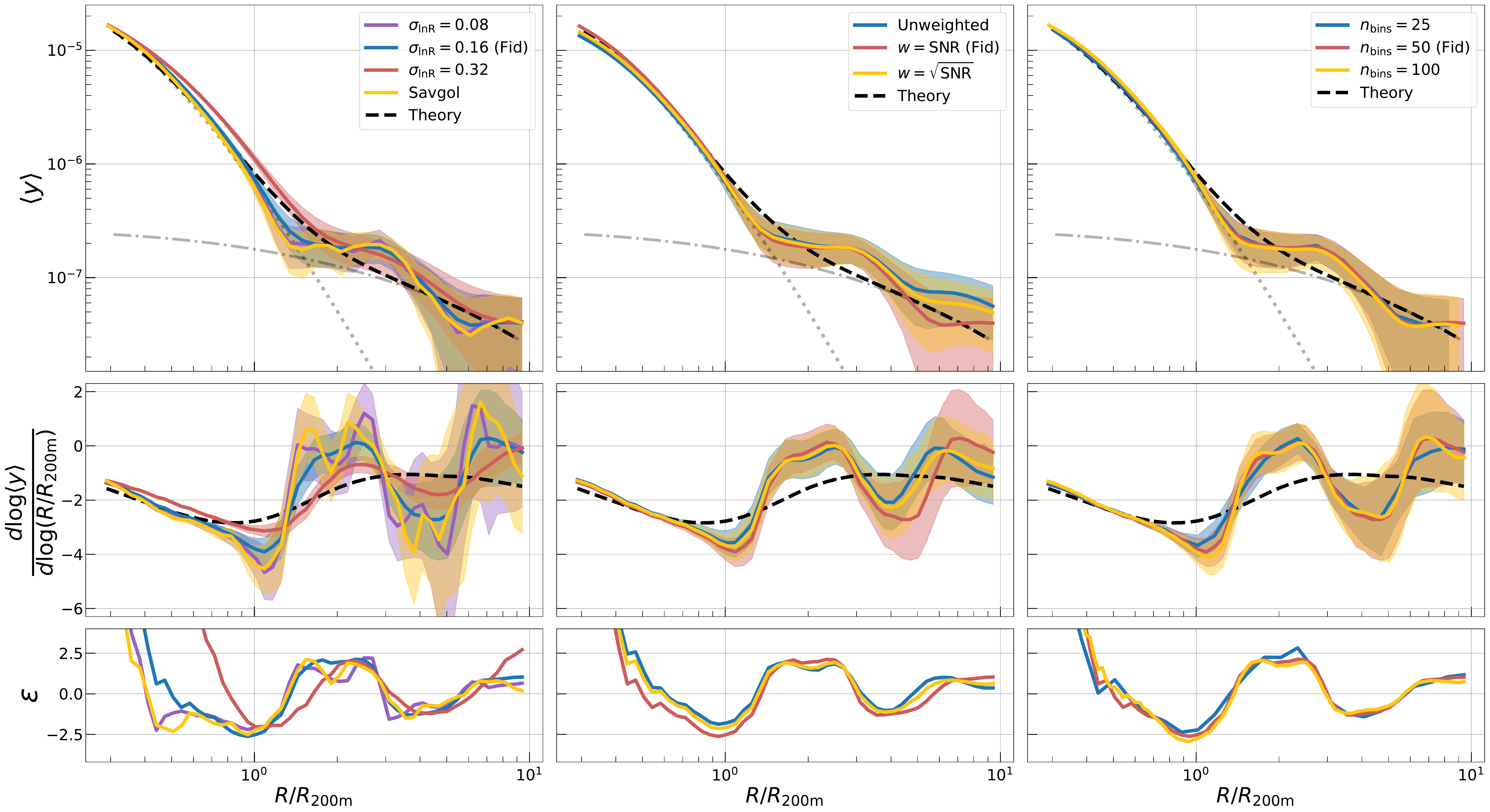}
    \caption{The response of our SPT-SZ results to changes in the analysis pipeline. From left to right we test: (i) smoothing kernels, (ii) weighting schemes, and (iii) bin widths. For each panel, we denote our fiducial choice as ``Fid'', and also show the theory curves, which are the same in every column. The results change slightly when increasing our Gaussian smoothing scale by a factor $2$, while the rest of the method variations have no relevant effects.}
    \label{fig:Robustness_Tests}
\end{figure*}

We have made certain methodological choices in our analysis, and here we test the sensitivity of our results to these choices. In Figure \ref{fig:Robustness_Tests}, we show the response of the SPT-SZ mean $y-$profile and log-derivative (previously shown in Figure \ref{fig:Detection_of_Shock}) to variations in the (i) smoothing kernels, (ii) weighting schemes, and; (iii) bin widths. We detail our findings below.

\textbf{Smoothing:} In general, all smoothing kernel choices still resolve the shock features at the same detection significance. When using a narrower Gaussian with width $\sigma_{\rm \ln R} = 0.08 < \dln R \approx 0.11$ ---which is effectively an unsmoothed case --- the log-derivative is noisier, particularly further out from the cluster center. Conversely, using a wider Gaussian with $\sigma_{\rm \ln R} = 0.32$ results in oversmoothing which both changes the profile significantly and dampens the deviations from the theory model as showcased in the fiducial SPT-SZ result (Figure \ref{fig:Detection_of_Shock}). The oversmoothing also shifts the profiles and log-derivative to larger scales. Using a Savitsky-Golay filter --- the original choice of Paper I --- still resolves the shock. Here we used a window length of 7 and a polynomial order of 3 for the input parameters of the Savitsky-Golay filter.

\textbf{Weights:} Our results are quite insensitive to the choices of weights used when stacking profiles. Using $w = \texttt{SNR}$ leads to the highest detection significance for both features, but the improvement is miniscule. Even in the unweighted case, we can resolve both shocks at close to the fiducial detection significance. Note that for SPT-SZ, $\texttt{SNR} \in [4.5, 42]$, and so $\sqrt{\texttt{SNR}} \in [2, 6.5]$ which is closer to the unweighted scenario given the smaller variation in weights. The mean profile beyond $R/\Rtwohm > 5$ does show some differences but these are not statistically significant. We do not consider the case of $w = \texttt{SNR}^2$ as the wide dynamic range of the resulting weights leads to a small fraction of clusters (10\%) dominating the final result. 

\textbf{Binning:} Finally, changing the bin widths by factors of 2 has no impact on the stacked profile and log-derivative. Note that we continue to use a Gaussian smoothing step with $\sigma_{\rm \ln R} = 0.16$ in all three cases.

\bsp	
\label{lastpage}
\end{document}